\documentclass[aps,prd,onecolumn,showpacs,eqsecnum,nofootinbib]{revtex4}
\usepackage{epsfig}
\usepackage{graphicx}
\usepackage{dcolumn}
\usepackage{amsmath}
\usepackage{enumerate}
\usepackage{subfigure}        

\def\beq{\begin{equation}}
\def\eeq{\end{equation}}

 \def\gtap{\mathrel{ \rlap{\raise 0.511ex \hbox{$>$}}{\lower 0.511ex
   \hbox{$\sim$}}}} 
\def\ltap{\mathrel{ \rlap{\raise 0.511ex
    \hbox{$<$}}{\lower 0.511ex \hbox{$\sim$}}}} 
\newcommand{\bea}{\begin{eqnarray}} \newcommand{\eea}{\end{eqnarray}}
\newcommand{\deltaunodue}{\mbox{$\Delta m_{21}^2 $}}
\newcommand{\deltaunotre}{\mbox{$\Delta m_{31}^2 $}}

\begin{document}

\vskip-6pt \hfill {IPPP/09/84} \\
\vskip-6pt \hfill {DCPT/09/168} \\
\vskip-6pt \hfill {MPP-2009-167} \\
\vskip-6pt \hfill{EURONU-WP6-09-10}\\
\title{On the improvement of the low energy neutrino factory}

\author{Enrique Fern\'andez Mart\'{\i}nez}
\email{enfmarti@mppmu.mpg.de}
\affiliation{Max-Planck-Institut f\"ur Physik (Werner-Heisenberg-Institut),
F\"ohringer Ring 6, 80805 M\"unchen, Germany}
\author{Tracey Li}
\email{t.c.li@durham.ac.uk}
\affiliation{IPPP, Department of Physics, Durham University, Durham, DH1 3LE, United Kingdom}
\author{Olga Mena} 
\email{omena@ific.uv.es}
\affiliation{Instituto de F\'{\i}sica Corpuscular, IFIC CSIC and
  Universidad de Valencia, Spain}
\author{Silvia Pascoli}
\email{silvia.pascoli@durham.ac.uk}
\affiliation{IPPP, Department of Physics, Durham University, Durham, DH1 3LE, United Kingdom}

\begin{abstract}
The low energy neutrino factory has been proposed as a very sensitive setup for future searches for CP violation and matter effects. Here we study how its performance is affected when the experimental specifications of the setup are varied. Most notably, we have considered the addition of the `platinum' $\nu_\mu \rightarrow \nu_{e}$ channel. We find that, whilst theoretically the extra channel provides very useful complementary information and helps to lift degeneracies, its practical usefulness is lost when considering realistic background levels. Conversely, an increase in statistics in the `golden' $\nu_e \rightarrow \nu_\mu$ channel and, to some extent, an improvement in the energy resolution, lead to an important increase in the performance of the facility, given the rich energy dependence of the `golden' channel at these energies. 
We show that a low energy neutrino factory with a baseline of 1300 km, muon energy of 4.5 GeV, and either a 20 kton totally active scintillating detector or 100 kton liquid argon detector, can have outstanding sensitivity to the neutrino oscillation parameters $\theta_{13}$, $\delta$ and the mass hierarchy. For our estimated exposure of $2.8\times10^{23}$ kton $\times$ decays per muon polarity, the low energy neutrino factory has sensitivity to $\theta_{13}$ and $\delta$ for $\sin^{2}(2\theta_{13}) > 10^{-4}$ and to the mass hierarchy for $\sin^{2}(2\theta_{13}) > 10^{-3}$.
\end{abstract}

\pacs{14.60.Pq}

\maketitle

\section{Introduction}\label{sec:intro}

Neutrino oscillations have been robustly established. The present data require two large ($\theta_{12}$ and $\theta_{23}$) angles and one small ($\theta_{13}$) angle in the neutrino mixing matrix, and at least two mass squared differences, $\Delta m_{ij}^{2} \equiv m_i^2 -m_j^2$ (where $m_{i}$'s are the neutrino masses), one driving the atmospheric ($\deltaunotre$) and the other one the solar ($\deltaunodue$) neutrino oscillations. The mixing angles $\theta_{12}$ and $\theta_{23}$ control the solar and the atmospheric neutrino oscillations, while $\theta_{13}$ is the angle which connects the atmospheric and solar neutrino regimes.

A global fit performed at the end of 2008~\cite{SchwetzFit} (see also~\cite{concha}) provides
the following $3\sigma$ allowed ranges for the atmospheric mixing
parameters: $|\Delta m_{31}^{2}| =(2.07 - 2.75)\times10^{-3}$~eV$^{2}$
and $0.36<\sin^2\theta_{23}<0.67$. The sign of $\Delta m_{31}^{2}$
($\mathrm{sign} (\Delta m^2_{31})$) cannot be determined from the existing data. The two possibilities, 
$\deltaunotre > 0$ or $\deltaunotre < 0$, correspond to two different 
types of neutrino mass ordering: normal hierarchy and inverted
hierarchy respectively. 
In addition, information on the octant of $\theta_{23}$, 
if $\sin^22\theta_{23} \neq 1$, is beyond the reach of present
experiments. 
The best fit values for the solar neutrino oscillation parameters are 
$\Delta m_{21}^{2}=7.65 \times 10^{-5}~{\rm eV^2}$ and
$\sin^2\theta_{12}=0.30$~\cite{SchwetzFit}. 
A non-zero value of $\theta_{13}$ is crucial to enable a measurement
of the CP violating phase $\delta$ and the mass hierarchy. 
A combined three-neutrino
oscillation analysis of the solar, atmospheric, reactor and
long-baseline neutrino data~\cite{SchwetzFit} constrains the third
mixing angle to be $\sin^2\theta_{13}\leq 0.056$ at the $3\sigma$
confidence level, with a best fit value of $0.01$. 
Different analyses undertaken in 2008 using the available solar data from the third phase of the Sudbury Neutrino Observatory (SNO-III) and recent data from KamLAND favour a non-zero value of $\sin^{2}\theta_{13}$ at $\sim 1\sigma$. Similarly, Super-Kamiokande data on atmospheric neutrinos~\cite{Fogli:2008jx} leads to a $\sim 2\sigma$ preference for $\sin^{2}\theta_{13}>0$. This second claim is, however, controversial~\cite{Maltoni:2008ka}.
2009 data from the MINOS experiment, studying the appearance channel $\nu_{\mu}\rightarrow\nu_{e}$, also favours a non-zero value of $\theta_{13}$ but with an even larger best fit, even more in conflict with the stringent upper bound that comes mainly from the CHOOZ reactor experiment of $\sin^2\theta_{13}\leq 0.056$.
A preliminary combination of all the data provides a $1\sigma$ range of $\sin^{2}\theta_{13}=0.02\pm0.01$~\cite{th13}. This hint for non-zero $\theta_{13}$ and the resulting tension among the different datasets will be probed by the forthcoming generation of accelerator \cite{t2k,nova} and reactor \cite{chooz2,dayabay,reno} experiments. If the hint for large $\theta_{13}$ is confirmed, the exciting possibility to search for leptonic CP violation, encoded in the phase $\delta$, and the ordering of neutrino masses \cite{Huber:2009cw}, will be open. However, these experiments lack the required sensitivity and a new generation of neutrino oscillation experiments is therefore needed for this task, and to explore even smaller values of $\theta_{13}$ if the present hint is not confirmed.

Future long-baseline experiments will require powerful machines and
extremely pure neutrino beams. Among these, neutrino factories~\cite{geer}, 
in which a neutrino beam is generated from muons decaying within 
the straight sections of a storage ring, have been shown to be
sensitive tools for studying neutrino oscillation
physics~\cite{geer,nf1,nf2,nf4,nf5,nf6,silver,study1physics,nf8,study2,yo,nf9,GeerMenaPascoli,hubernew,minenf,lownf2}. The neutrino factory exploits the golden signature of the
\emph{wrong-sign} muon~\cite{geer,nf1} events, i.e. muons 
with opposite sign to the muons stored in the neutrino factory. 
Wrong-sign muons ($\mu^-$) result from $\nu_{e} \rightarrow \nu_{\mu}$
oscillations (if $\mu^+$ are stored), and can be used to measure the
mixing angle $\theta_{13}$, determine the neutrino mass hierarchy, and
search for CP violation in the neutrino sector. In addition to the 
\emph{wrong-sign} muon signal, there will also be
\emph{right-sign} muon events. These events come from the
disappearance muon neutrino channel, $\bar{\nu}_{\mu} \rightarrow
\bar{\nu}_{\mu}$ ($\nu_{\mu} \rightarrow \nu_{\mu}$), if positive
(negative) muons are stored. 
The discrimination of the \emph{wrong} and \emph{right} sign muons 
requires the identification of charged current (CC) muon neutrino
interactions, and the measurement of the sign of the produced muon. 
If the interacting neutrinos have energies of more than a few GeV, 
standard neutrino detector technology, based on large magnetised
 sampling calorimeters, can be used to measure wrong-sign muons 
with high efficiency and very low backgrounds. 
This has been shown to work for neutrino factories with energies of
about 20 GeV or greater~\cite{nf5,ISS-Detector Report,study1physics}.
 
Lower energy neutrino factories~\cite{GeerMenaPascoli,lownf2}, which
store muons with energies $<10$~GeV, exploit a fully active
calorimeter within a magnet, a detector technology which ensures the
detection of lower energy muons. The possibility of a low energy neutrino factory with non-magnetic detectors has also been explored in~\cite{nonmagnetic}, although we have not considered it here. A neutrino factory with muon energies
of about 4 GeV has been shown to enable very precise measurements of
the neutrino mixing parameters~\cite{GeerMenaPascoli,lownf2}. Electron
charge identification also becomes possible in a low energy 
neutrino factory equipped with a magnetised totally active
scintillating detector (TASD)~\cite{ISS-Detector Report}. 
Therefore, in addition to the wrong and right-sign
\emph{muons}, there will also be wrong and right-sign \emph{electrons} 
 from the appearance channel (the \emph{platinum} channel), $\bar{\nu}_{\mu}\rightarrow \bar{\nu}_e$, and the disappearance
channel, ${\nu}_{e}\rightarrow \nu_{e}$, 
for positive muons stored in the decay ring~\footnote{Distinguishing the
electron signature from the neutral current events will represent 
a very difficult task for the magnetised calorimeter technology.}.
These \emph{platinum} channels,
which are the T-conjugates of the golden channels, could provide a  
possible way of resolving the problem of degenerate solutions
~\cite{FL96,MN01,BMWdeg,deg}. It is well known that even a very 
precise measurement of the appearance probability for neutrinos 
and antineutrinos at a fixed $L/E$ allows for different solutions 
of $(\theta_{13}, \mathrm{sign} (\Delta m^2_{31}), \delta)$, 
severely weakening the sensitivity to these parameters. 
Many strategies have been advocated to resolve this issue which 
in general involve another detector~\cite{MN97,BMW02off,SN1,
  twodetect,SN2,T2kk}, the combination with another experiment~
\cite{otherexp1,BMW02,HLW02,MNP03,otherexp,mp2,HMS05,Choubey,yo,
  nf9,huber2,lastmine} and/or the addition of new
channels~\cite{silver,CPTchannels}. 

We will consider the impact of the addition of the platinum 
channels in both a low statistics and high statistics scenario. 
We will show that if the platinum channel is accessible with negligible backgrounds (by which we mean a charge misidentification rate of $<10^{-2}$, and that fewer than $10^{-2}$ of all neutral current events are wrongly counted as signal events - see Section~\ref{subsec:plat} for more details), in the case of low statistics, its addition would greatly increase the potential of the low energy neutrino factory. However for the high statistics scenario, the platinum channel has much less of an effect - higher statistics combined with complementary information from the different energies of a broad beam alone are sufficient to resolve degeneracies and maximise the performance of the setup. Unfortunately, once realistic backgrounds from misidentified pions and the challenging electron charge identification are taken into account, the addition of the platinum channel does not provide any significant improvement. 

The structure of the paper is as follows: in
Section~\ref{sec:physics} we discuss in detail the physics reach of
the proposed setup, which exploits the wrong and right sign muon and
electron signals. Based on the reasons described in Section~\ref{sec:physics}, the decaying muon
statistics assumed here is higher than in previous studies. The assumptions for the detection efficiencies and energy resolution of
the detector have also been modestly improved, based on the simulated performance of the NO$\nu$A TASD detector \cite{nova}. These assumptions provide a more competitive setup with
respect to our previous studies~\cite{GeerMenaPascoli,lownf2}. We
perform detailed numerical simulations and discuss the sensitivity of
the low energy neutrino factory to the mixing angle $\theta_{13}$, to
the CP violating phase $\delta$, to the neutrino mass hierarchy, to the octant of the atmospheric mixing angle $\theta_{23}$ and to deviations from maximal atmospheric mixing
as a function of the energy resolution of the detector and the number
of muon decays per year (with and without the addition of the 
platinum channels). 
In Section~\ref{sec:LAr} we introduce our preliminary studies 
of a magnetised 100 kton liquid argon (LAr) detector, 
comparing its performance to that of the TASD and other near term and
future long-baseline neutrino facilities. Finally, in Section~\ref{sec:conc}, we draw our conclusions.

\section{Physics reach: optimisation of the experimental setup}\label{sec:physics}

In this section we present the results from numerical simulations of the low energy neutrino factory. We have used the GLoBES software package~\cite{globes} to simulate several experimental configurations which will be described in the following subsections. These have led us to an optimised setup, which we use unless otherwise specified, defined by the following: the baseline is 1300 km, corresponding to the Fermilab to DUSEL distance. For the beam we consider a muon energy of 4.5 GeV with $1.4\times10^{21}$ useful muon decays per year per polarity, running for ten years. This flux is larger than the $2.5\times10^{20}$ useful muon decays per year per polarity usually considered for the International Design Study (IDS) neutrino factory~\cite{ISS-Detector Report} for three reasons. Firstly, the IDS neutrino factory distributes the beam to two baselines whilst only one baseline is required for the low energy neutrino factory - therefore a factor of two is gained. Moreover, $1\times10^{7}$ operational seconds per year were assumed in~\cite{ISS-Detector Report} whilst we consider that $2\times10^{7}$ operational seconds per year should be achievable. However, in order to perform an equal comparison with other facilities, $1\times10^{7}$ seconds per year were assumed in the comparison plots of Section~\ref{sec:LAr}. Finally, the extra factor of 1.4 arises from a re-optimisation of the accelerator complex for the lower muon energy required with respect to the IDS design~\cite{reoptimisedacc}. 

For the detector we assume a totally active scintillating detector (TASD) with a fiducial mass of 20 kton, energy threshold of 0.5 GeV, energy resolution of $10\%$ with 19 variable-width bins, efficiency for $\mu^{\pm}$ detection of $73\%$ below 1 GeV and $94\%$ above, and a background level of $10^{-3}$ on the $\nu_{e}\rightarrow\nu_{\mu}$ ($\bar{\nu}_{e}\rightarrow\bar{\nu}_{\mu}$) and $\nu_{\mu}\rightarrow\nu_{\mu}$ ($\bar{\nu}_{\mu}\rightarrow\bar{\nu}_{\mu}$) channels. We assume that the background to each channel arises predominantly from charge misidentification and neutral current events, modeling the background to each channel as a constant fraction of the rates of the wrong-sign and neutral current channels. For systematics, we use $2\%$ on both the signal and background, assuming the errors to be uncorrelated.

For the platinum ($\nu_{\mu}\rightarrow\nu_{e}$ and $\bar{\nu}_{\mu}\rightarrow\bar{\nu}_{e}$) channels, identifying the charges of the electrons at the energies involved will be very challenging. In addition, the pion background is very difficult to separate from the electron samples as compared to the muon samples. Preliminary estimations~\cite{private} suggest that the efficiency for $e^{\pm}$ detection could be $37\%$ below 1 GeV and $47\%$ above, and that the backgrounds (we make similar assumptions for the background sources as for the muon channels) could be reduced to a level of $10^{-2}$. Although it is uncertain as to whether this detector performance can actually be achieved - further simulations are required - we will regard these numbers as an optimistic estimate for the electron detection capabilities of a TASD.

We assume the same oscillation parameters as in~\cite{baseline spec}: $\sin^{2}\theta_{12}=0.3$, $\theta_{23}=\pi/4$, $\Delta m_{21}^{2}=8.0\times10^{-5}$ eV$^{2}$, and $|\Delta m_{31}^{2}|=2.5\times10^{-3}$ eV$^{2}$ with a $10\%$ uncertainty on the atmospheric parameters, $4\%$ uncertainty on the solar parameters, and $2\%$ uncertainty on the matter density. In all our simulations we have used the exact oscillation probabilities, taking into account matter effects, and have marginalised over all parameters.

The most significant alteration relative to the previous setup ~\cite{GeerMenaPascoli,lownf2} consists of the addition of the $\nu_{\mu}\rightarrow\nu_{e}$ and $\bar{\nu}_{\mu}\rightarrow\bar{\nu}_{e}$ channels. These \emph{platinum} channels are the T-conjugates of the golden channels ($\nu_{e}\rightarrow\nu_{\mu}$, $\bar{\nu}_{e}\rightarrow\bar{\nu}_{\mu}$), and we will investigate the power of this combination to reduce the degeneracies in the $\theta_{13}$, $\delta$ and sign($\Delta m_{31}^{2}$) parameter space. It has been shown that typically, the elimination of these degenerate solutions may require additional information from a second baseline and detector~\cite{MN97,silver,BMW02off,SN1,twodetect,SN2,T2kk}, or from a complementary experiment~\cite{otherexp1,BMW02,HLW02,MNP03,otherexp,mp2,HMS05,Choubey,yo,nf9,huber2,lastmine}. 

We try instead to exploit the ability of the TASD to detect and identify the charge of $e^{-}$ and $e^{+}$, which gives access to the platinum channel. The probability for this channel, to leading order in the small quantities $\theta_{13}$, $\alpha = \Delta m_{21}^{2}/\Delta m_{31}^{2}$ and $EA/\Delta m_{31}^{2}$ (where $A$ = $\sqrt{2}G_{\mathrm{F}}n_{e}$ is the matter potential and $n_{e}$ is the electron number density), is identical to that for the golden channel~\cite{nf5} with the interchange of $\delta\rightarrow -\delta$ and is shown below

\begin{subequations}
\begin{eqnarray}
P_{\mu e} &=& s_{213}^{2}s_{23}^{2}\sin^{2}(\frac{\Delta m_{31}^{2}L}{4E}-\frac{AL}{2})\label{eq:atm}\\
         &+& \alpha s_{213}s_{212}s_{223}\frac{\Delta m_{31}^{2}}{2EA}\sin(\frac{AL}{2})\sin(\frac{\Delta m_{31}^{2}L}{4E}-\frac{AL}{2})
\cos(\frac{\Delta m_{31}^{2}L}{4E}-\delta)\label{eq:CP}\\
         &+& \alpha^{2}c_{23}^{2}s_{212}^{2}\left(\frac{\Delta m_{31}^{2}}{2EA}\right)^{2}\sin^{2}(\frac{AL}{2}),\label{eq:solar}
\end{eqnarray}
\end{subequations}

\begin{subequations}
\begin{eqnarray}
P_{e\mu} &=& s_{213}^{2}s_{23}^{2}\sin^{2}(\frac{\Delta m_{31}^{2}L}{4E}-\frac{AL}{2})\label{eq:atm2}\\
         &+& \alpha s_{213}s_{212}s_{223}\frac{\Delta m_{31}^{2}}{2EA}\sin(\frac{AL}{2})\sin(\frac{\Delta m_{31}^{2}L}{4E}-\frac{AL}{2})
\cos(\frac{\Delta m_{31}^{2}L}{4E}+\delta)\label{eq:CP2}\\
         &+& \alpha^{2}c_{23}^{2}s_{212}^{2}\left(\frac{\Delta m_{31}^{2}}{2EA}\right)^{2}\sin^{2}(\frac{AL}{2}).\label{eq:solar2}
\end{eqnarray}
\end{subequations} We use a notation where $s_{ij} = \sin\theta_{ij}$, $s_{2ij} = \sin(2\theta_{ij})$, $c_{ij} = \cos\theta_{ij}$, $c_{2ij} = \cos(2\theta_{ij})$, $E$ is the neutrino energy and $L$ is the baseline. The first line of each probability, subequations (\ref{eq:atm}) and (\ref{eq:atm2}), is the \emph{atmospheric term} which is quadratic in $\sin(2\theta_{13})$ and will be dominant in the scenario that $\theta_{13}$ is large ($\sin^{2}(2\theta_{13})\gtrsim10^{-2}$), and at high energies. The atmospheric term provides sensitivity to $\theta_{13}$, the mass hierarchy, and is sensitive to the octant of $\theta_{23}$. The second line, subequations (\ref{eq:CP}) and (\ref{eq:CP2}), is the \emph{CP term} which is linear in $\sin(2\theta_{13})$ and dominates for intermediate values of $\theta_{13}$ if $\delta\neq$ $0$ or $\pi$. The dependence on $\delta$ enters via the oscillatory cosine term which can take either a positive or negative sign depending on the value of the phase. This can lead to constructive or destructive interference between the atmospheric and CP terms, meaning that sensitivities to $\theta_{13}$ and the mass hierarchy are strongly dependent on the value of $\delta$. Due to the inverse dependence on energy, the CP term becomes most visible at lower energies; therefore it is important to have access to the second oscillation maximum to establish if CP is violated. Thus a lower energy is desirable to enable a clean measurement of $\delta$, whereas a higher energy and, especially, a long-baseline, guarantees sensitivity to the mass hierarchy. The low energy neutrino factory is unique in having a surprising degree of sensitivity to the mass hierarchy in spite of its low energy (as we show in Section~\ref{subsec:TASDresults}) due to its broad spectrum that includes energies beyond the first oscillation peak, thus enabling complementary information to be obtained to resolve degeneracies. The third line, subequations (\ref{eq:solar}) and (\ref{eq:solar2}), is the \emph{solar term} which is independent of $\theta_{13}$, $\delta$ and the mass hierarchy, and is dominant in the case that $\theta_{13}$ is very small ($\sin^{2}(2\theta_{13})\lesssim10^{-4}$). In this regime, measurements will be extremely challenging and a high energy neutrino factory may be the only option~\cite{nf5,ISS-Detector Report,study1physics}.

If we consider the fact that the probability for the CP-conjugated golden channel, $\bar{\nu}_{e}\rightarrow\bar{\nu}_{\mu}$, takes a similar form to that of the golden channel but with the substitutions $\delta\rightarrow -\delta$ and $A\rightarrow -A$ and that the CPT-conjugated golden channel is identical to the golden channel, with the exchange of $A\rightarrow -A$, we can understand the complementarity of these four channels: each of the channels has a different dependence on the parameters $\theta_{13}$, $\delta$ and sign($\Delta m_{31}^{2}$) and so degenerate solutions are present at \emph{different} points in the parameter space for each of the channels~\cite{CPTchannels}. Thus the degenerate solutions from one channel can be eliminated by the information from another channel.

We will mention briefly that the ability of the TASD to detect electrons also enables measurement of the $\nu_{e}$ ($\bar{\nu}_{e}$) disappearance channel whose probability is
\begin{eqnarray}
P_{\nu_{e}\rightarrow\nu_{e}}&\approx& 1-s_{213}^{2}\sin^{2}\left(\frac{\Delta m_{31}^{2} L}{4E}-\frac{AL}{2}\right).
\end{eqnarray} However, as this channel is CP-invariant and has only a weak dependence on the mass hierarchy, it is expected and has been verified in our study that its addition does not provide any significant improvement.

In the rest of this section we now show the impact of our improved statistics and energy resolution, and of the addition of the platinum channels. 

\subsection{Energy resolution}\label{subsec:stats}

\begin{figure}[!here]
     \subfigure[~$\theta_{13} = 1^{\circ}$]{
          \includegraphics[width=8cm,height=6cm]{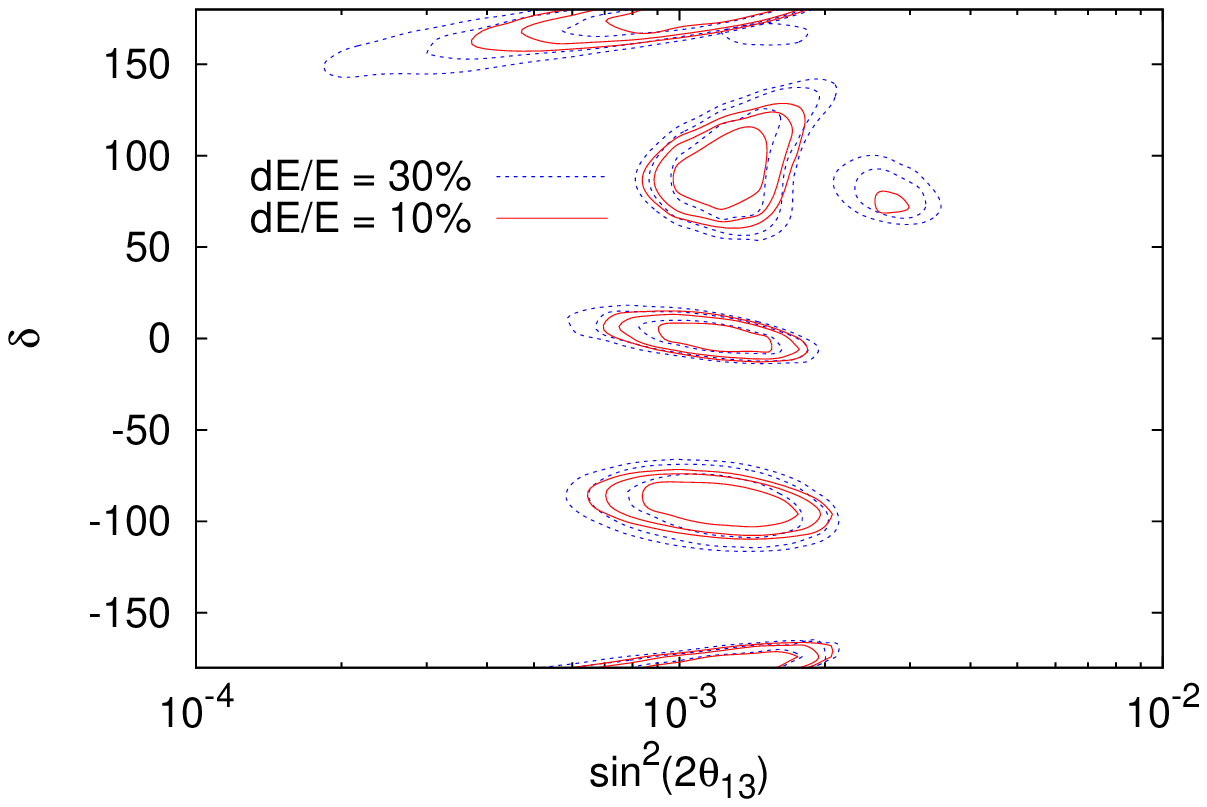}}
     \hspace{.3in}
     \subfigure[~$\theta_{13} = 5^{\circ}$]{
          \includegraphics[width=8cm,height=6cm]{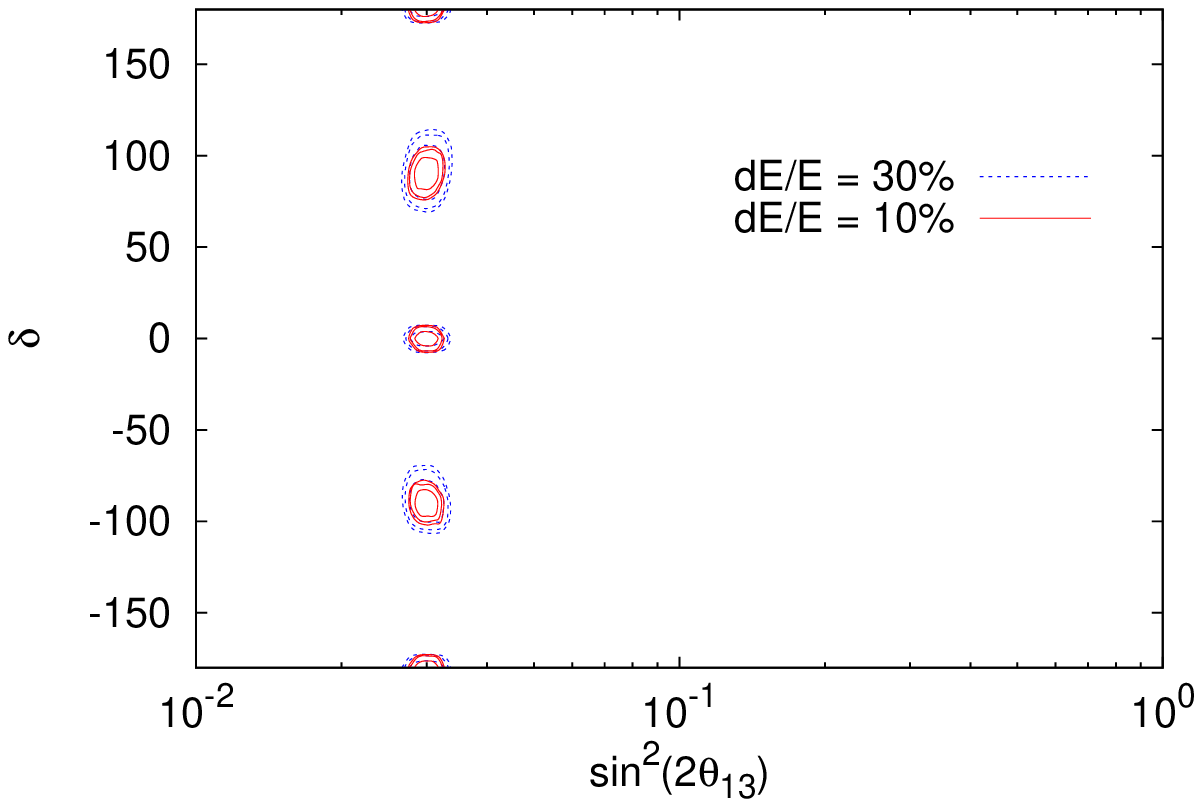}}\\
     \caption{Comparing an energy resolution of $dE/E = 30\%$ (dotted blue lines) and $10\%$ (solid red lines): $68\%$, $90\%$ and $95\%$ confidence level contours in the $\sin^{2}(2\theta_{13})-\delta$ plane for true values of $\delta=-180^{\circ},-90^{\circ}, 0^{\circ}$ and $90^{\circ}$ and a) $\theta_{13} = 1^{\circ}$, b) $\theta_{13} = 5^{\circ}$.}
\label{fig:res}
\end{figure}

Firstly we illustrate how our more optimistic estimate of $10\%$ for the energy resolution, $dE/E$, (with 19 variable width bins) improves the performance of our setup. In Fig.~\ref{fig:res} we show how the new resolution improves upon the old value of $30\%$ (with nine variable width bins) for $\theta_{13}=1^{\circ}$ and $5^{\circ}$. We observe that in addition to the significant increase in sensitivity to $\theta_{13}$ and $\delta$, the hierarchy degeneracy is now almost completely eliminated even for small values of $\theta_{13}$. This is possible because at these low energies, the oscillation probability displays a rich dependence on the oscillation parameters as a function of the neutrino energy. In particular, the hierarchy can be determined from the position of the first oscillation maximum, and the value of $\delta$ from both the first and second oscillation maxima. Hence if the neutrino energy can be measured with sufficient precision to enable the oscillation spectrum to be accurately reconstructed, a significant improvement in the sensitivities to the oscillation parameters can be achieved.

\subsection{Inclusion of the platinum channel}\label{subsec:plat}

\begin{figure}[!here]
     \subfigure[~$5.0\times10^{20}$ decays per year, zero background]{
          \includegraphics[width=8cm,height=6cm]{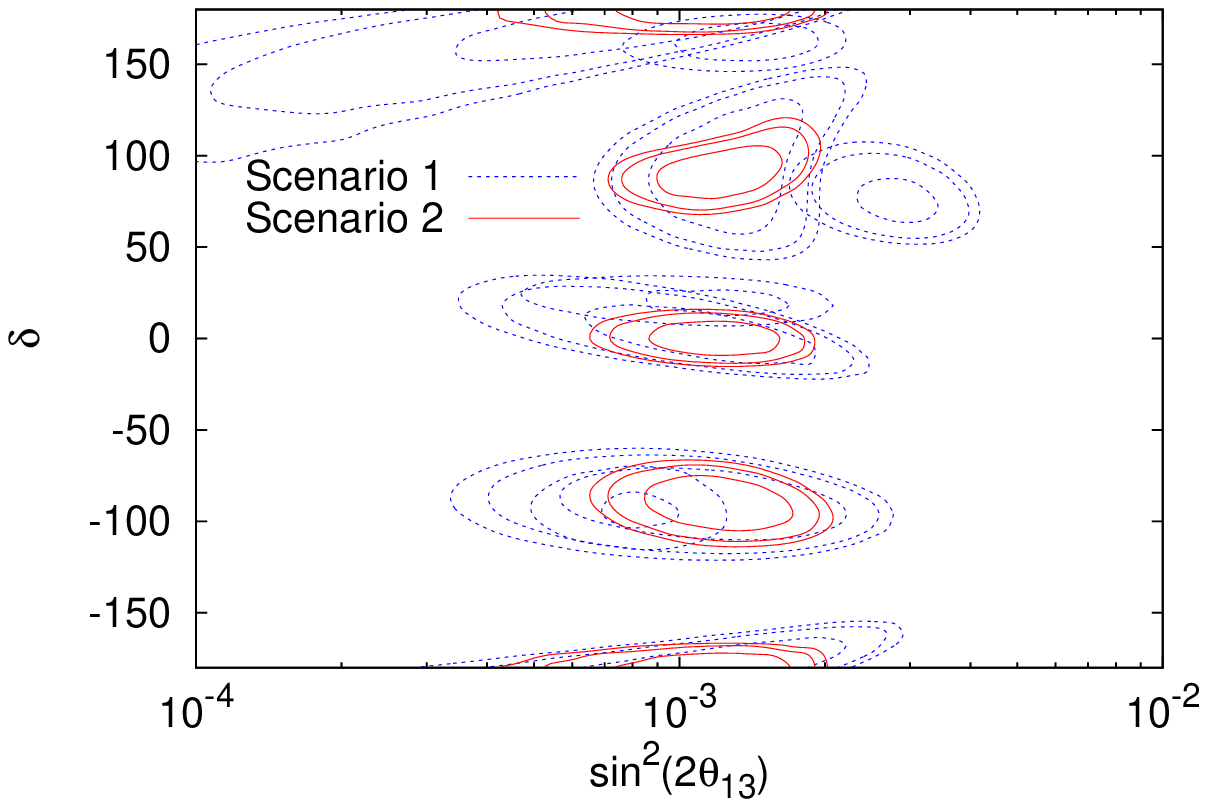}}
     \hspace{.3in}
     \subfigure[~$1.4\times10^{21}$ decays per year, zero background]{
          \includegraphics[width=8cm,height=6cm]{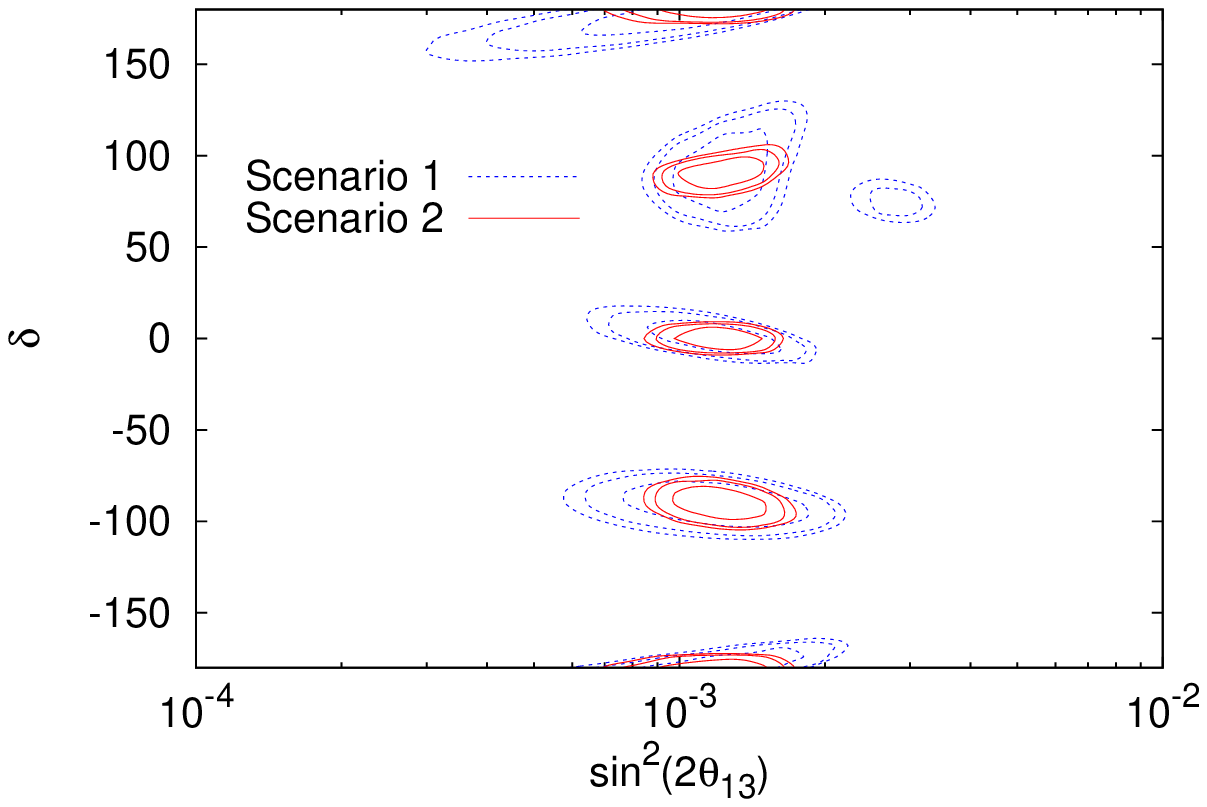}}\\
     \vspace{.3in}
     \subfigure[~$5.0\times10^{20}$ decays per year, $10^{-2}$ background]{
          \includegraphics[width=8cm,height=6cm]{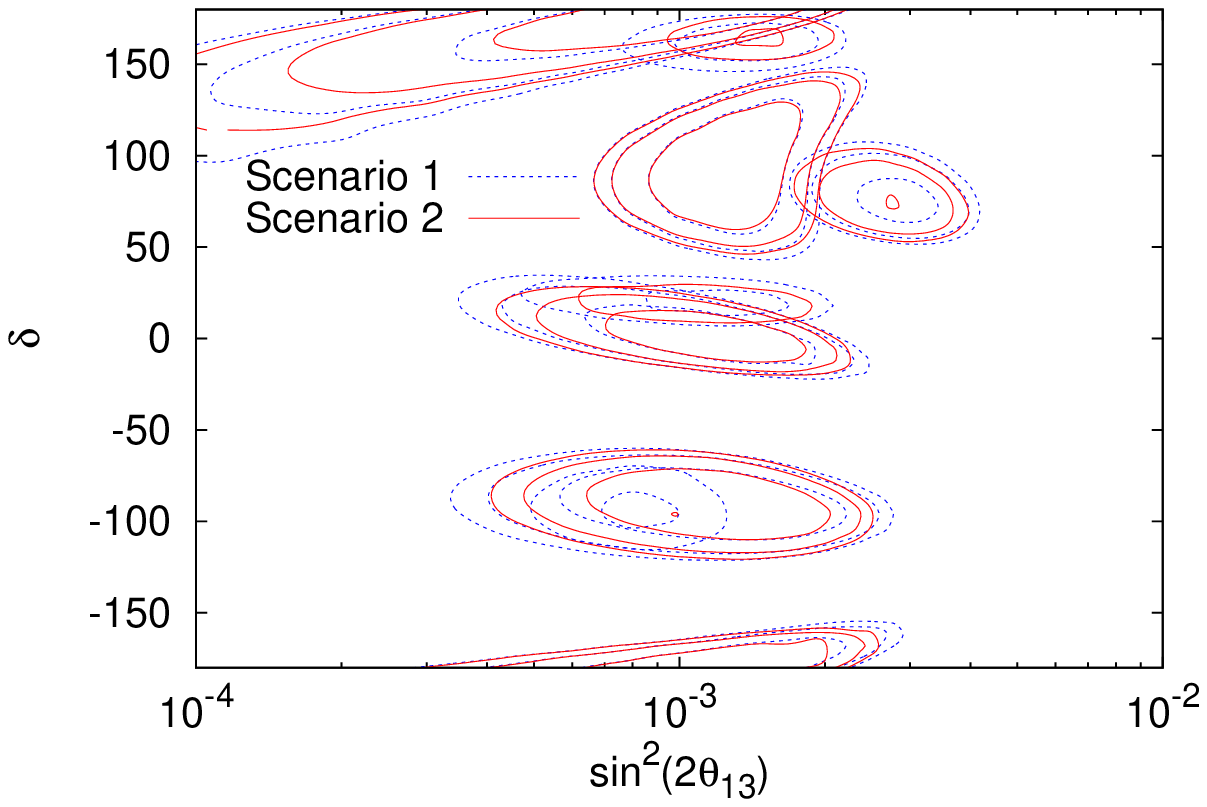}}
     \hspace{.3in}
     \subfigure[~$1.4\times10^{21}$ decays per year, $10^{-2}$ background]{
          \includegraphics[width=8cm,height=6cm]{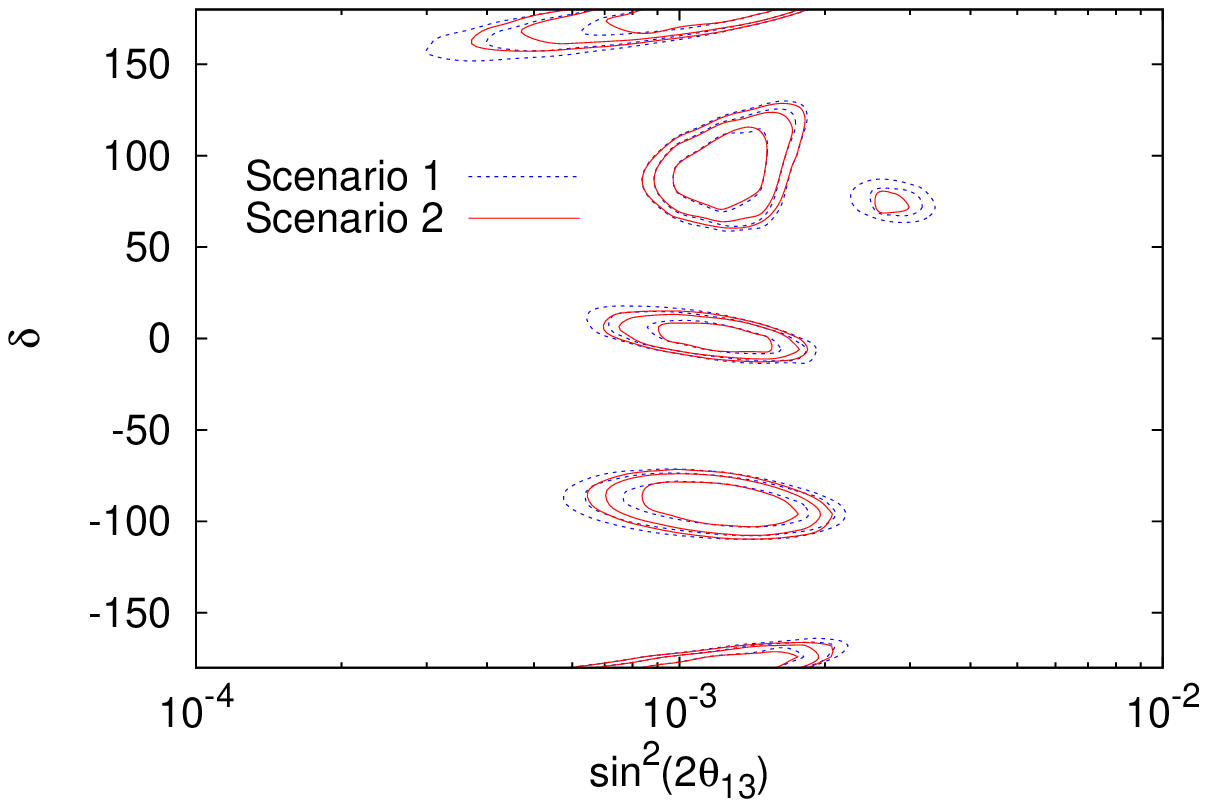}}\\
     \caption{Comparison of Scenario 1 ($\nu_{\mu}$ ($\bar{\nu}_{\mu}$) appearance and disappearance only - dotted blue lines), and Scenario 2 ($\nu_{e}$ ($\bar{\nu}_{e}$) appearance included - solid red lines) when using $5.0\times10^{20}$ $\mu^{\pm}$ decays per year (left) or $1.4\times10^{21}$ decays per year (right), and a background of zero (top row) or $10^{-2}$ (bottom row) on the $\nu_{e}$ ($\bar{\nu}_{e}$) channels: $68\%$, $90\%$ and $95\%$ confidence level contours in the $\sin^{2}(2\theta_{13})-\delta$ plane, for $\delta=-180^{\circ},-90^{\circ}, 0^{\circ}$ and $90^{\circ}$ and $\theta_{13} = 1^{\circ}$.}
\label{fig:bg}
\end{figure}

We define \emph{Scenario 1} to be the one in which only $\mu^{\pm}$ detection is possible, giving us access to only the $\nu_{\mu}$ and $\bar{\nu}_{\mu}$ appearance and disappearance channels. In \emph{Scenario 2} it is also possible to detect $e^{\pm}$ and hence exploit the additional information from the $\nu_{e}$ and $\bar{\nu}_{e}$ appearance channels. To illustrate the impact of the addition of the platinum channel to our setup, in Fig.~\ref{fig:bg} we compare the sensitivities of the two scenarios when using a muon decay rate of $5.0\times10^{20}$ (left column) and $1.4\times10^{21}$ (right column) per year, varying the background level of the $\nu_{e}$ ($\bar{\nu}_{e}$) appearance channel from a hypothetical zero (top row) to $10^{-2}$ (bottom row). By this we mean that we have simulated a charge misidentification rate such that $10^{-2}$ of all $e^{-}$ ($e^{+}$) are wrongly identified as $e^{+}$ ($e^{-}$), and that $10^{-2}$ of all $\nu_{e}$ ($\bar{\nu}_{e}$) neutral current events are wrongly counted as signal events.

In the case of the lower statistics, we observe that the addition of the platinum channel with zero background produces a drastic improvement in sensitivity to all parameters. For a background of $10^{-2}$ the improvement is much smaller but can still help to alleviate the hierarchy degeneracy (see~\cite{CPTchannels}). At higher backgrounds we find that this gain is lost. In the case of the high statistics, we already observe a smaller improvement for zero background, which becomes insignificant at a background level of $10^{-2}$. Thus we conclude that since the estimated background on the $\nu_{e}$ ($\bar{\nu}_{e}$) appearance channels will be at best $\sim 10^{-2}$, the platinum channel
could help in the measurement of the mass hierarchy if statistics are limited to $5.0\times10^{20}$ useful muon decays per year, whereas it will be almost irrelevant for the higher statistics scenario. An increase in statistics in the golden channel thus provides a much larger improvement in the facility performance than the addition of the platinum channel if background levels below $10^{-2}$ cannot be achieved.

\subsection{Sensitivity to $\theta_{13}$, CP violation and mass ordering}\label{subsec:TASDresults}

\begin{figure}[!here]
     \subfigure[~$\theta_{13}$ discovery potential]{
          \includegraphics[scale=0.5]{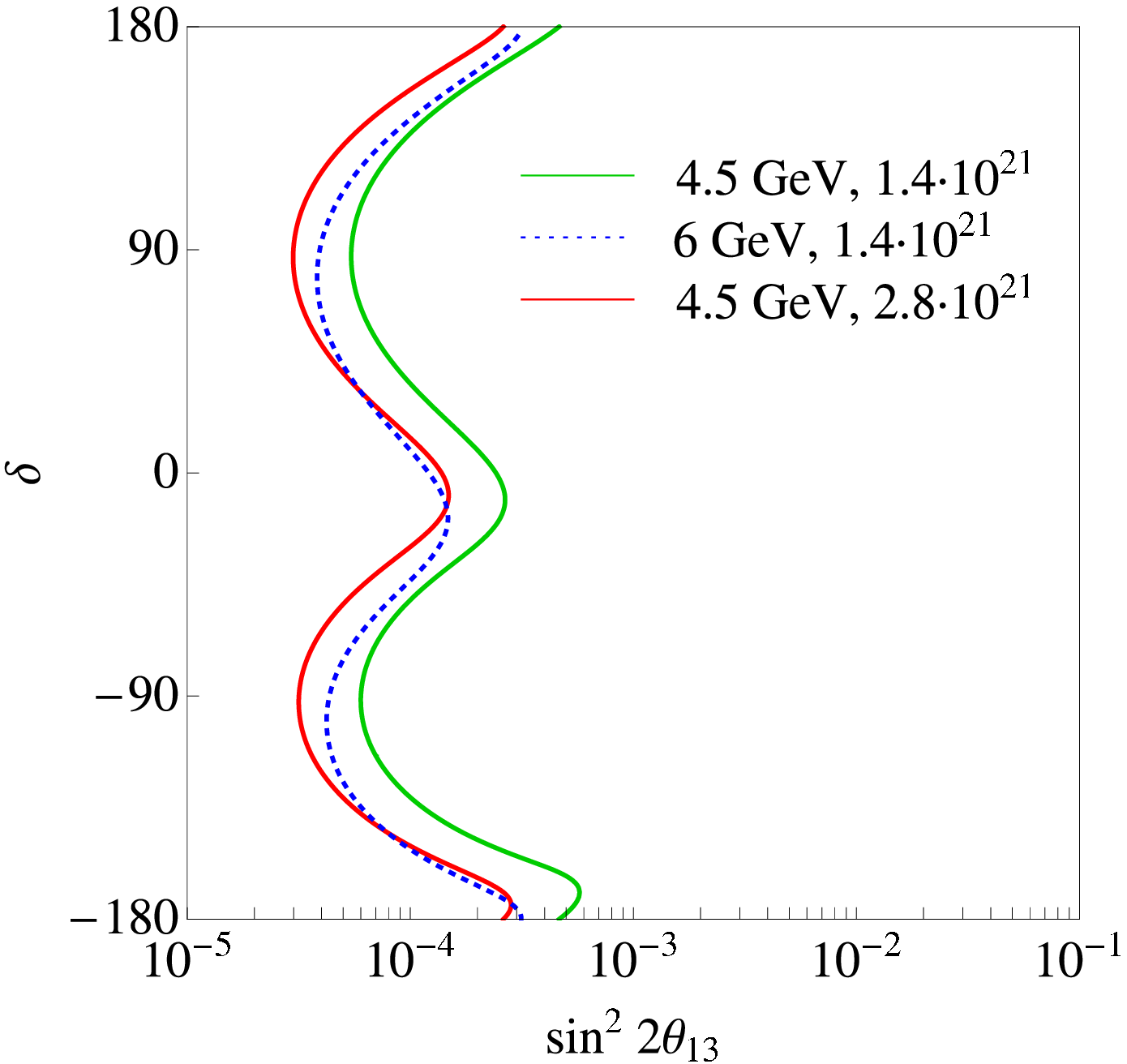}}
     \hspace{.3in}
     \subfigure[~CP discovery potential]{
          \includegraphics[scale=0.5]{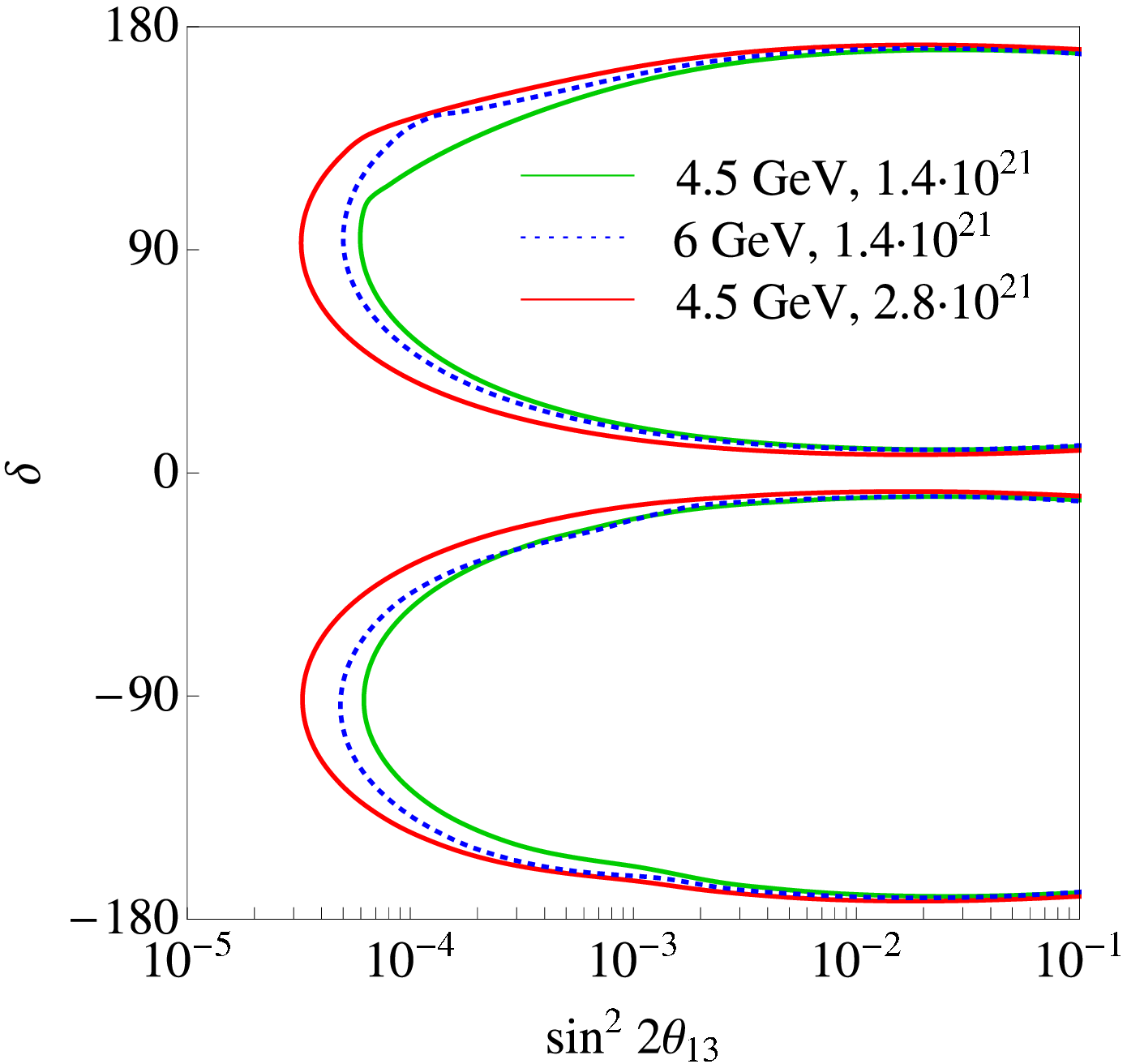}}\\
     \vspace{.3in}
     \subfigure[~Hierarchy sensitivity (normal hierarchy)]{
          \includegraphics[scale=0.5]{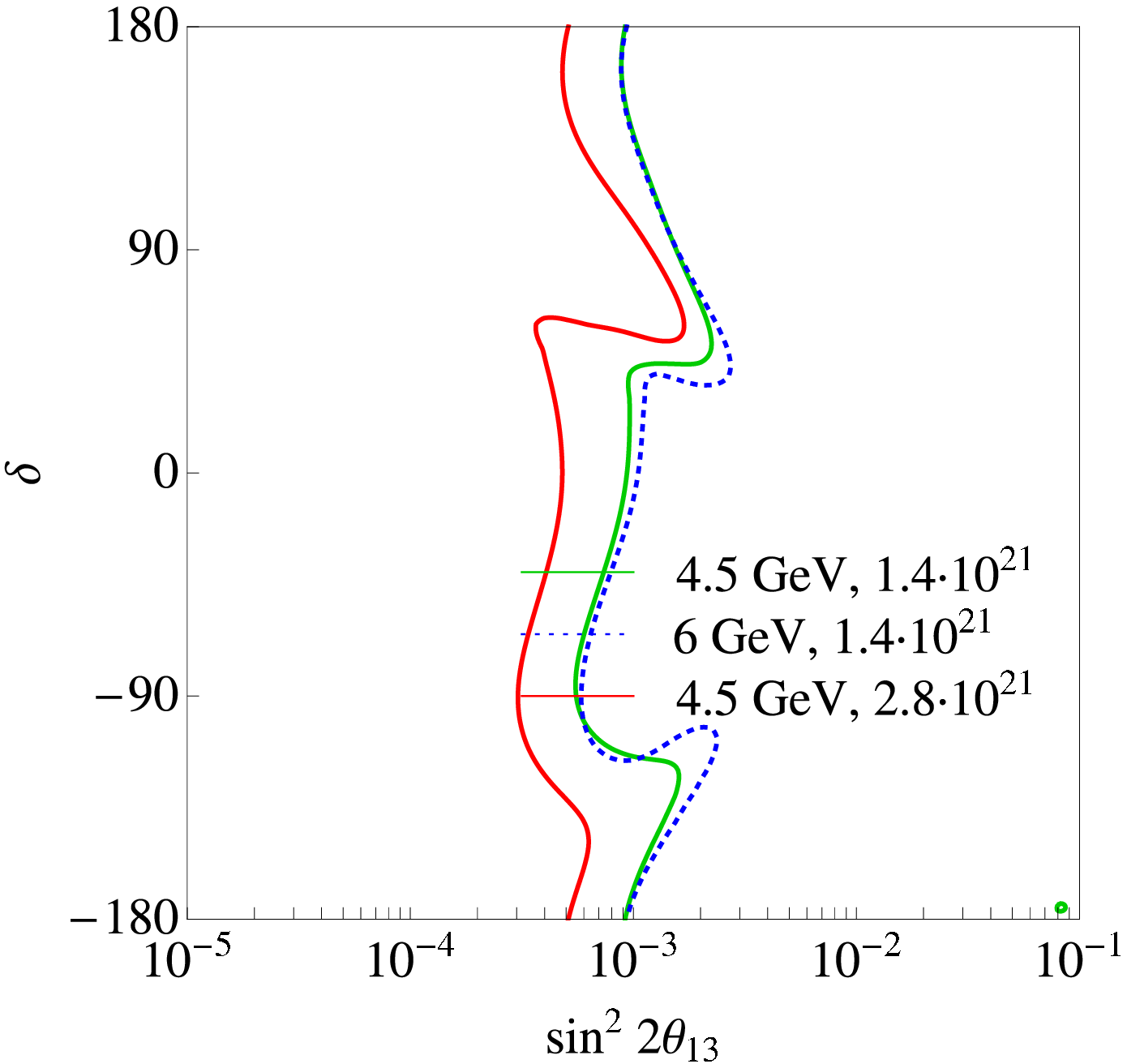}}
     \hspace{.3in}
     \subfigure[~Hierarchy sensitivity (inverted hierarchy)]{
          \includegraphics[scale=0.5]{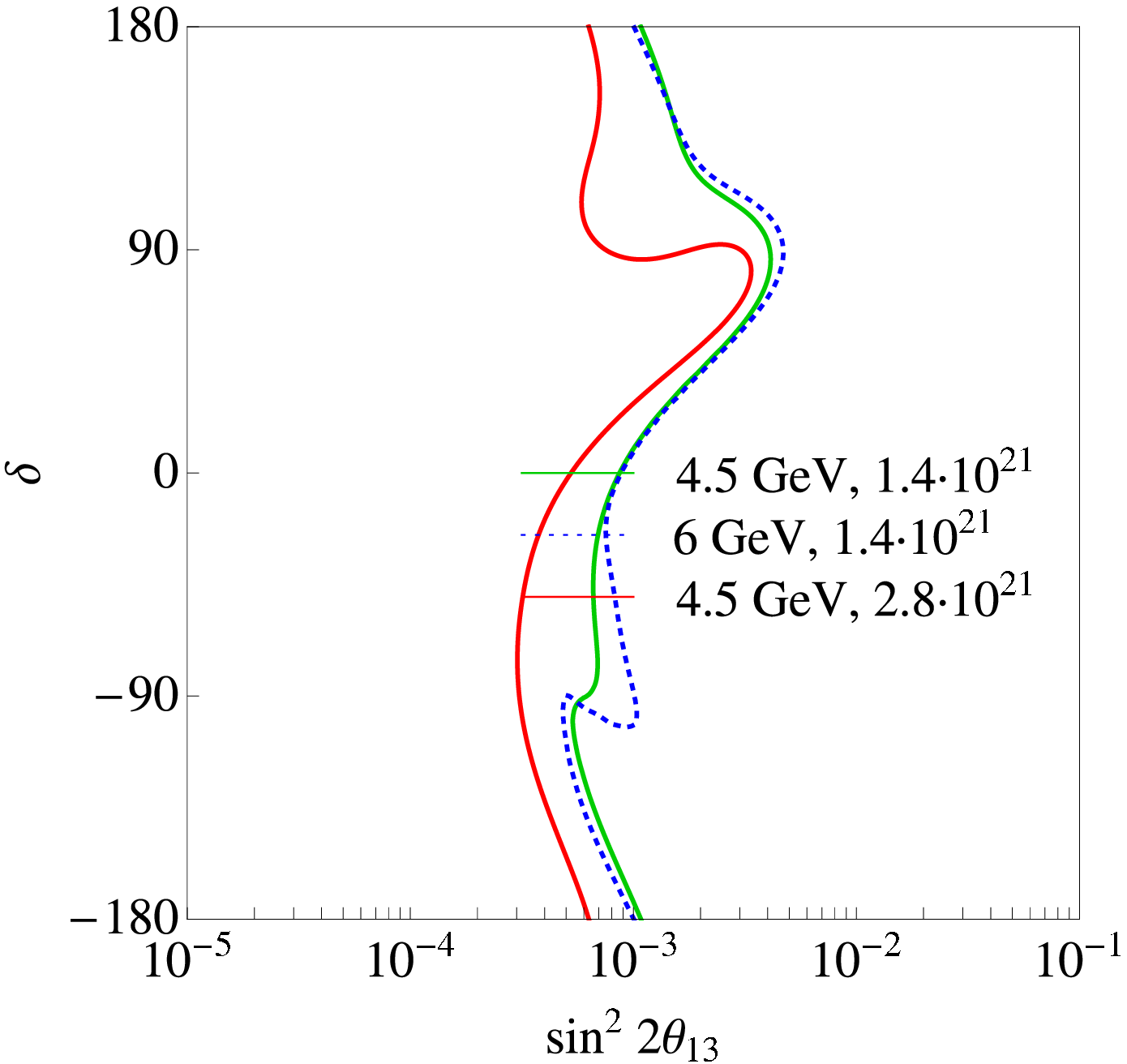}}\\
     \caption{$3\sigma$ confidence level contours in the $\sin^{2}(2\theta_{13})-\delta$ plane for a) $\theta_{13}$ discovery potential, b) CP discovery potential, c) hierarchy sensitivity (for true normal hierarchy), d) hierarchy sensitivity (for true inverted hierarchy).}
\label{fig:TASD_sens}
\end{figure}

\begin{figure}
     \subfigure[~Sensitivity to $\theta_{23}\neq45^{\circ}$]{
          \includegraphics[scale=0.5]{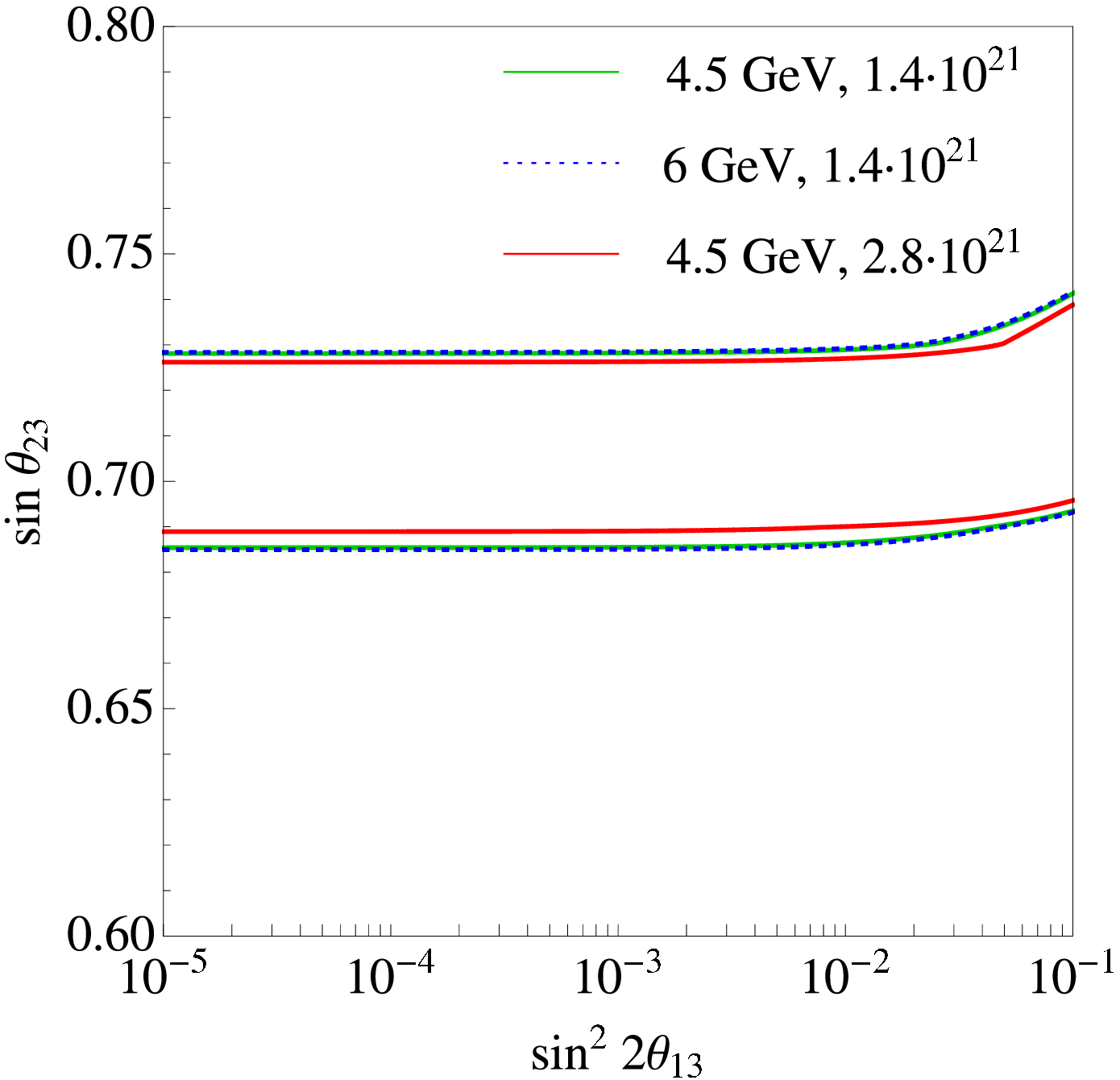}}
     \hspace{.3in}
     \subfigure[~Sensitivity to the $\theta_{23}$ octant]{
          \includegraphics[scale=0.5]{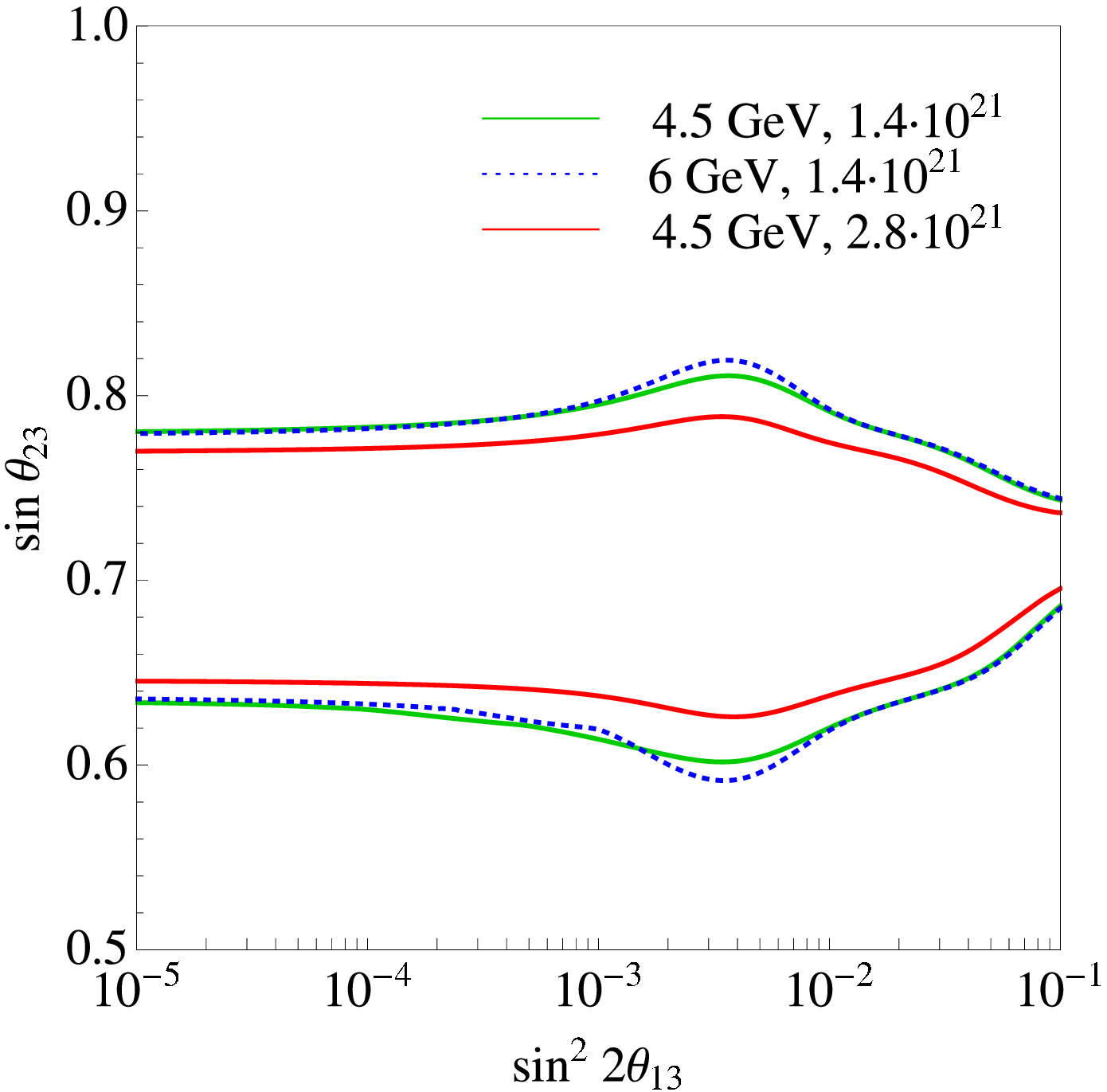}}\\
     \caption{$3\sigma$ allowed regions in the $\sin^{2}(2\theta_{13})-\sin\theta_{23}$ plane for a) potential to exclude $\theta_{23} = 45^{\circ}$, b) sensitivity to the $\theta_{23}$ octant.}
\label{fig:TASD_th23}
\end{figure}

Here we present the results of our optimisation studies, in terms of $3\sigma$ $\theta_{13}$ discovery potential, CP discovery potential, and sensitivity to the mass hierarchy in the $\sin^{2}(2\theta_{13})-\delta$ plane (Fig.~\ref{fig:TASD_sens}). In addition we also consider the $3\sigma$ sensitivity to $\theta_{23}$ in the $\sin^{2}(2\theta_{13})-\sin\theta_{23}$ plane, both in terms of the ability to exclude a maximal value of $\theta_{23}$ (Fig.~\ref{fig:TASD_th23}a) and to identify the octant of $\theta_{23}$ (Fig.~\ref{fig:TASD_th23}b). The results from our optimised setup described in Section~\ref{sec:physics} are shown by the solid green lines. We have also considered a setup where only the statistics are altered, to $2.8\times10^{21}$ decays per year (solid red lines), and a setup where only the muon energy is increased to 6.0 GeV (dashed blue lines). From this we demonstrate that for all the observables considered, doubling the flux is always preferable to an increase in energy.

For $\theta_{13}$ discovery potential, CP discovery potential and $\theta_{23}$ sensitivity we only show the results for a normal hierarchy, having verified that similar results are obtained for an inverted hierarchy. We have assumed in Fig.~\ref{fig:TASD_th23} ($\theta_{23}$ sensitivity) a value of $\delta=90^{\circ}$ although we have also studied other values of $\delta$ and find no strong dependence on the CP phase, since sensitivity to $\theta_{23}$ is mainly obtained from terms with no dependence on $\delta$ in the oscillation probabilities discussed in Section~\ref{sec:physics}. For the exclusion of $\theta_{23}=45^{\circ}$, an upward curve is seen for large $\theta_{13}$. This can be understood 
because the addition of a large $\theta_{13}$ to the $\nu_\mu$ disappearance probability introduces an asymmetry in $\theta_{23}$ that shifts the contours to larger values (see
eq.~(1) and Fig.~8 of Ref.~\cite{Stef}).

We note that this setup has remarkable sensitivity to $\theta_{13}$ and $\delta$ for values of $\sin^{2}(2\theta_{13}) > 10^{-4}$, and
that its sensitivity to the mass hierarchy is an order of magnitude better that that of other proposed experiments exploiting the same baseline e.g. the wide-band beam experiment in~\cite{WBB}. We can attribute these qualities to the unique combination of high statistics and good background rejection coupled with an intermediate baseline, allowing for a clean measurement of the CP phase whilst also allowing for the mass hierarchy to be determined for $\sin^{2}(2\theta_{13}) > 10^{-3}$.

We have also explored how the precision with which $\theta_{13}$, $\delta$ and the deviation from maximal $\theta_{23}$ could eventually be measured at the low energy neutrino factory, varies as a function of exposure (detector mass $\times$ decays) per polarity. Our standard setup corresponds to 20 kton $\times$ $1.4\times10^{21}$ decays/ year $\times$ 10 years  $=2.8\times10^{23}$ kton $\times$ decays per polarity. We find that if the mixing angle $\theta_{13}$ turns out to be large, the unknown leptonic mixing parameters could be measured with unprecedented precision at a future low energy neutrino factory for sufficiently high exposures. The gain in precision is much less pronounced for values larger than $6 \times 10^{23}$ kton $\times$ decays per polarity, hence it may not be worth trying to increase the exposure beyond this value. 

\begin{figure}
     \centering
     \subfigure[~Sensitivity to $\theta_{13}$, including systematic errors and backgrounds]{
          \includegraphics[scale=0.5]{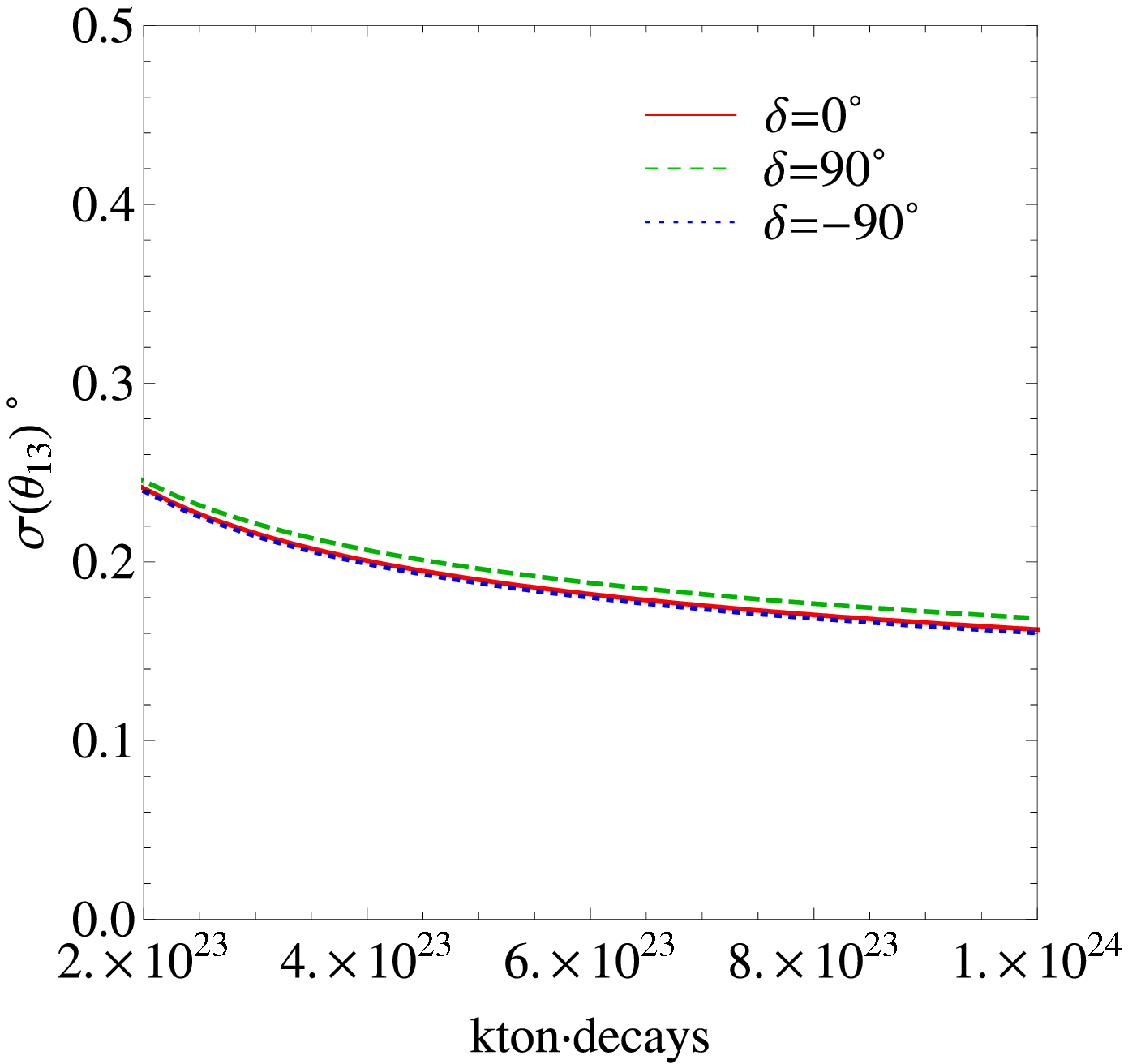}}
     \hspace{.3in}
     \subfigure[~Sensitivity to $\theta_{13}$, no systematic errors and backgrounds]{
          \includegraphics[scale=0.5]{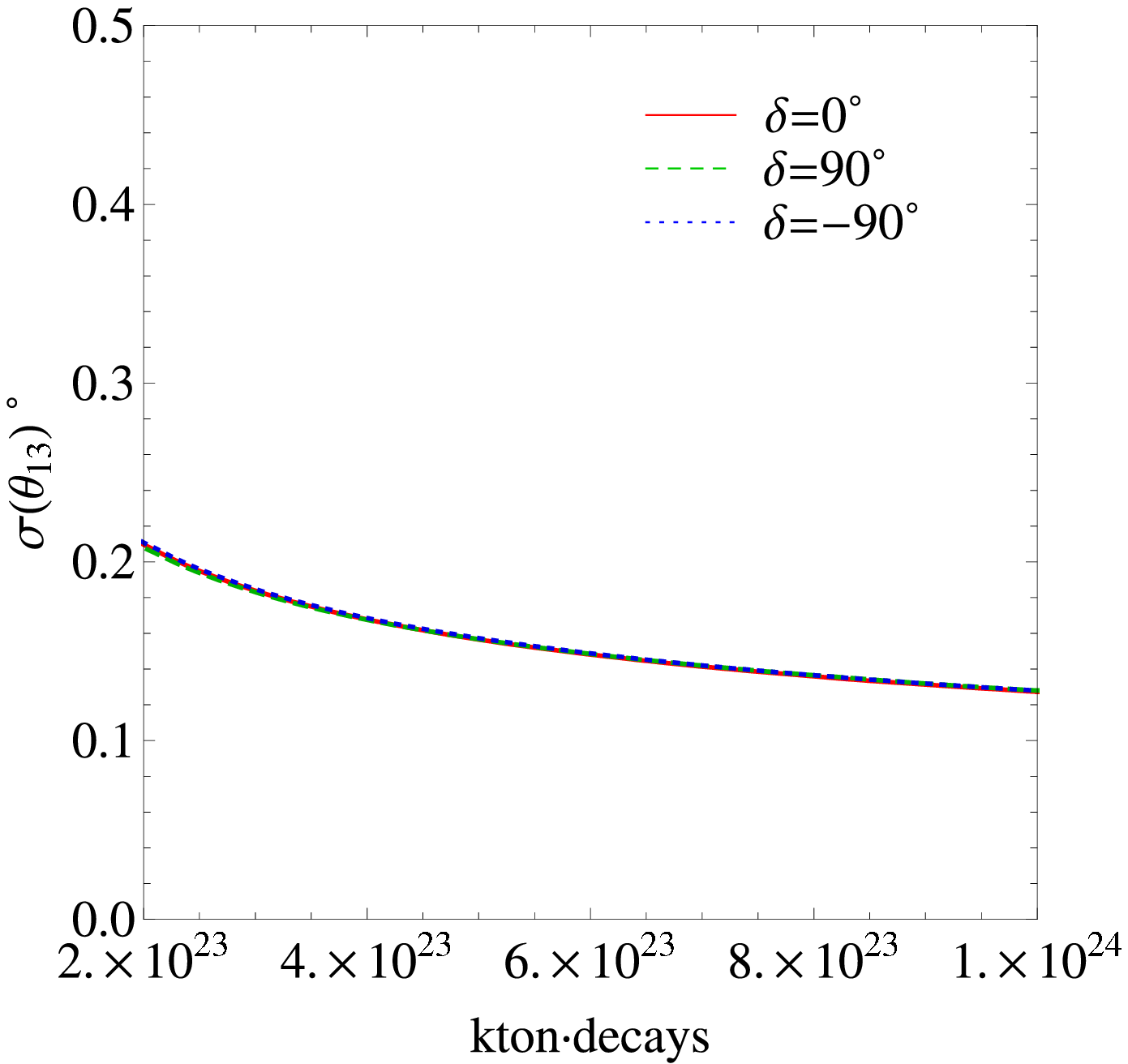}}\\
     \caption{$1\sigma$ error in the measurement of the $\theta_{13}$ mixing angle for a simulated value of $\theta_{13}=5^\circ$ and different values of the CP violating phase $\delta$ when a) including systematic errors and backgrounds, b) no systematic errors and backgrounds are included.}
\label{fig:prectheta}
\end{figure}

Fig.~\ref{fig:prectheta}a shows the $1\sigma$ error expected in the measurement of the mixing angle $\theta_{13}$ at a future low energy neutrino factory as a function of the exposure (in kton $\times$ decays) per polarity, assuming that nature has chosen $\theta_{13}= 5^\circ$. The dependence of these results on the value of the CP violating phase is very mild. The $1\sigma$ error in the extraction of $\theta_{13}$ when no backgrounds and no systematic errors are included in the analysis is illustrated in Fig.~\ref{fig:prectheta}b. Comparing the two panels we observe that non-zero systematics and backgrounds effectively halve the exposure.

\begin{figure}
     \centering
     \subfigure[~Sensitivity to $\delta$, including systematic errors and backgrounds]{
          \includegraphics[scale=0.5]{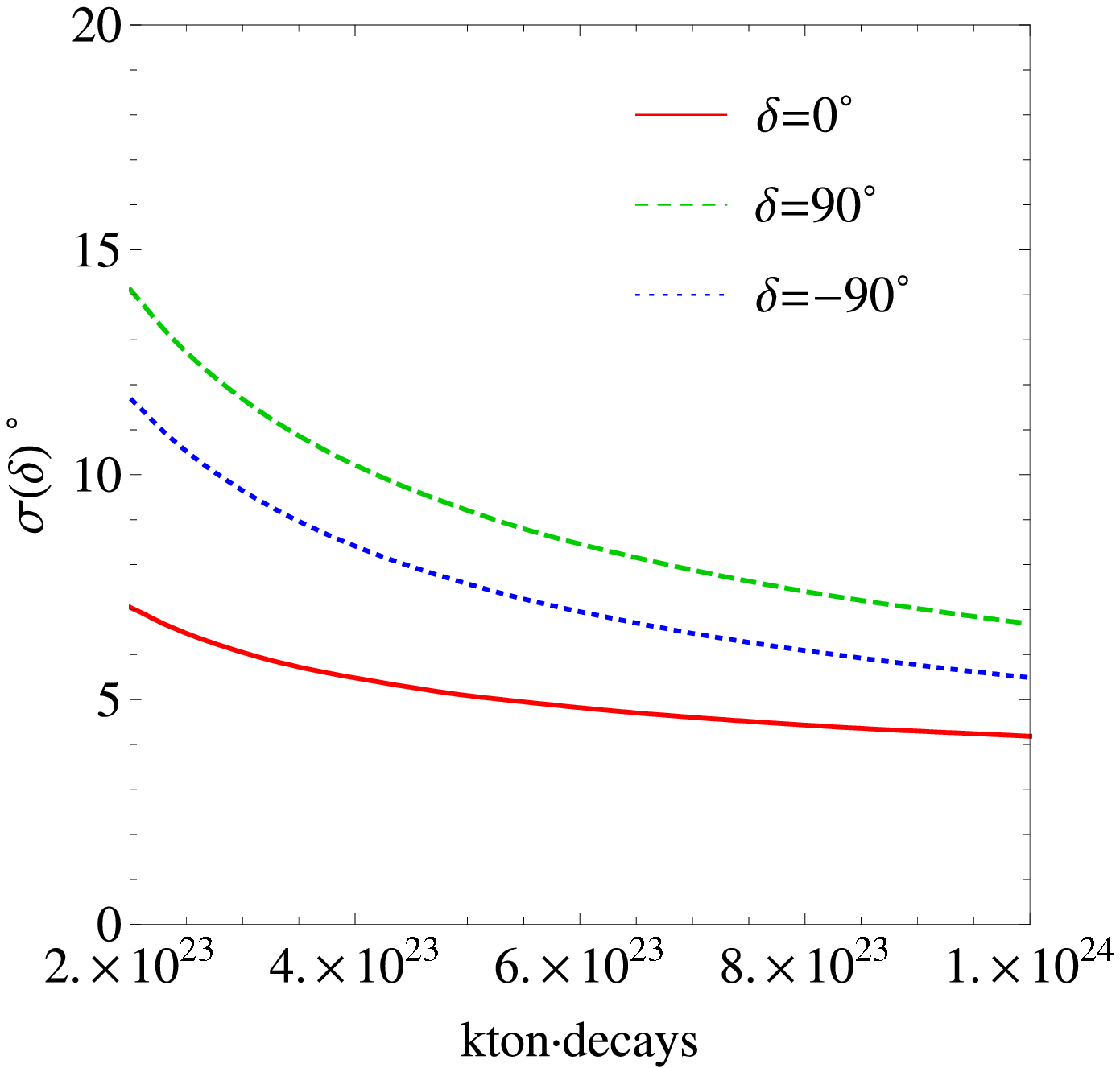}}
     \hspace{.3in}
     \subfigure[~Sensitivity to $\delta$, no systematic errors and backgrounds]{
          \includegraphics[scale=0.5]{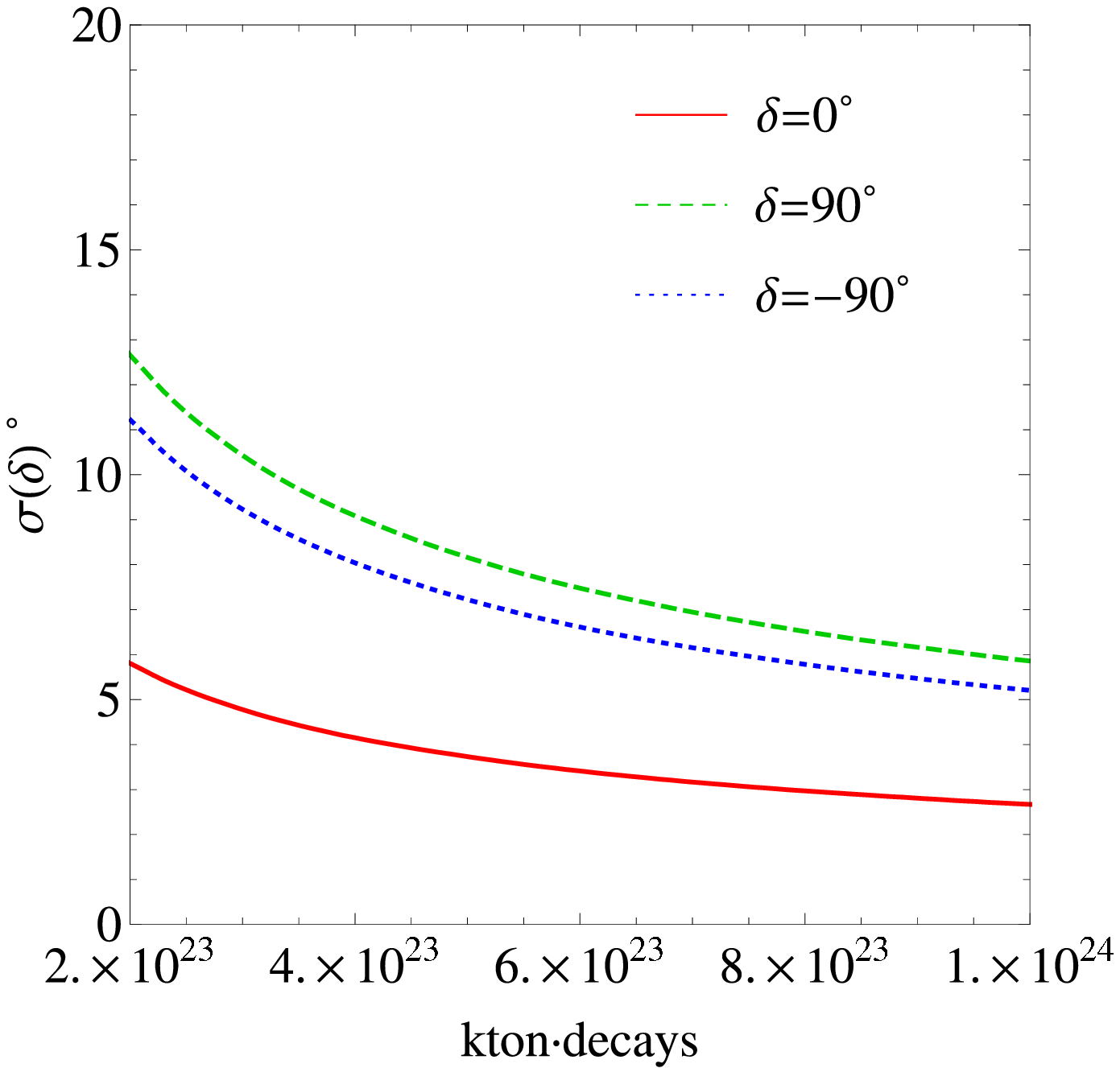}}\\
     \caption{$1\sigma$ error in the measurement of the CP violating phase $\delta$ for a simulated value of $\theta_{13}=5^\circ$ and different values of the CP violating phase $\delta$ when a) including systematic errors and backgrounds, b) no systematic errors and backgrounds are included.}
\label{fig:precdelta}
\end{figure}

Fig.~\ref{fig:precdelta}a shows the $1\sigma$ error expected in the measurement of the CP phase $\delta$ as a function of the exposure for a simulated value of $\theta_{13}= 5^\circ$, for different values of the CP violating phase $\delta$. The results are highly dependent on the value of the CP violating phase, as expected. For $\delta=90^\circ$, there are strong correlations with $\theta_{13}$, as can be seen from Fig.~\ref{fig:bg}, and therefore the error in the measurement of the CP violating phase is larger. The $1\sigma$ error in the extraction of $\delta$ when no backgrounds and no systematic errors are included in the
analysis is illustrated in Fig.~\ref{fig:precdelta}b. Switching off systematic errors and backgrounds has a larger impact for the
$\delta=0^\circ$ case, again effectively halving the exposure, since correlations among $\delta$ and $\theta_{13}$ are negligible when $\delta=0^\circ$ and the precision is more limited by the background and systematic errors instead.

\begin{figure}
     \centering
     \subfigure[~Sensitivity to $\theta_{23}\neq45^{\circ}$, including systematic errors and backgrounds]{
          \includegraphics[scale=0.5]{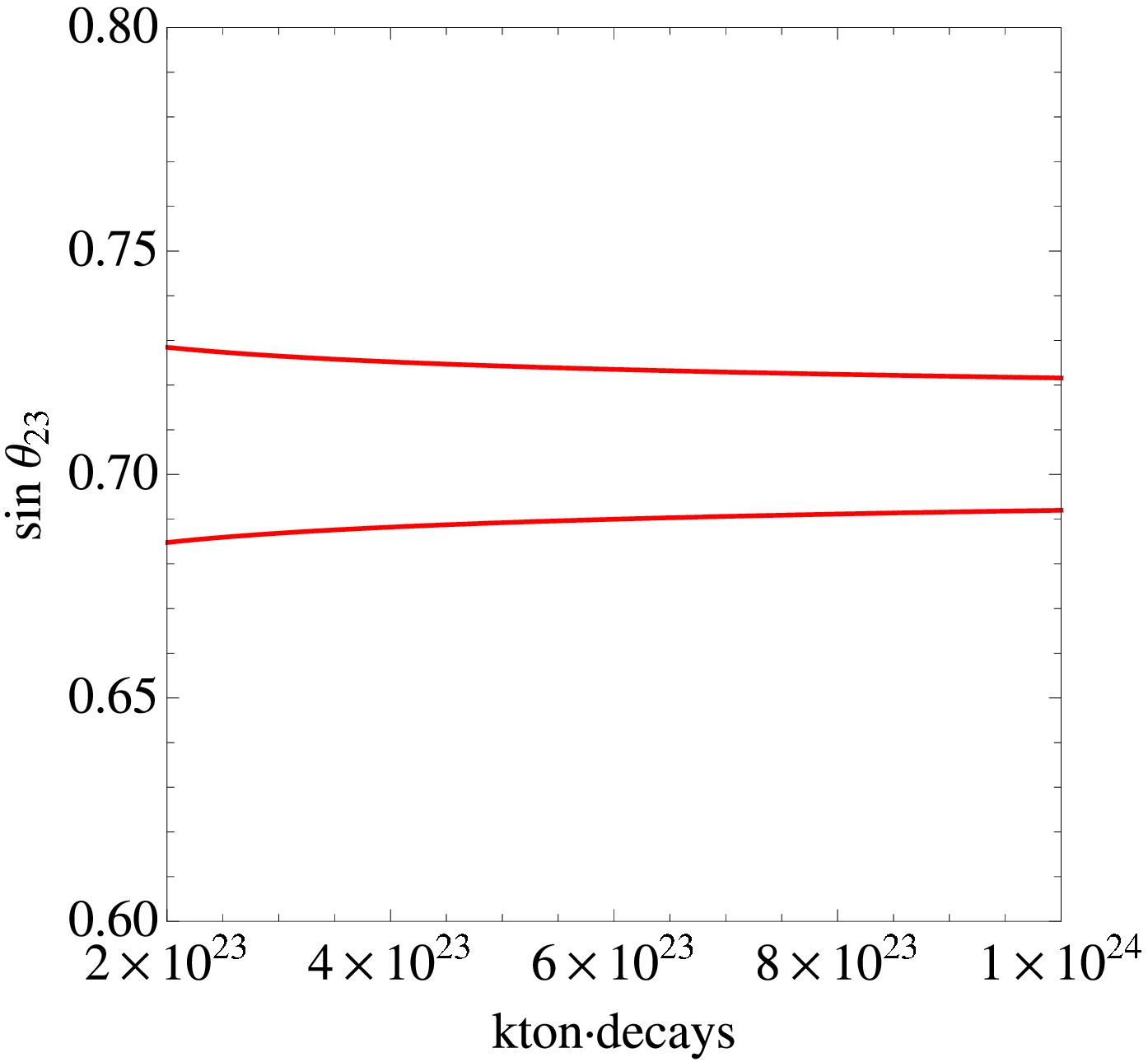}}
     \hspace{.3in}
     \subfigure[~Sensitivity to $\theta_{23}\neq45^{\circ}$, no systematic errors and backgrounds]{
          \includegraphics[scale=0.5]{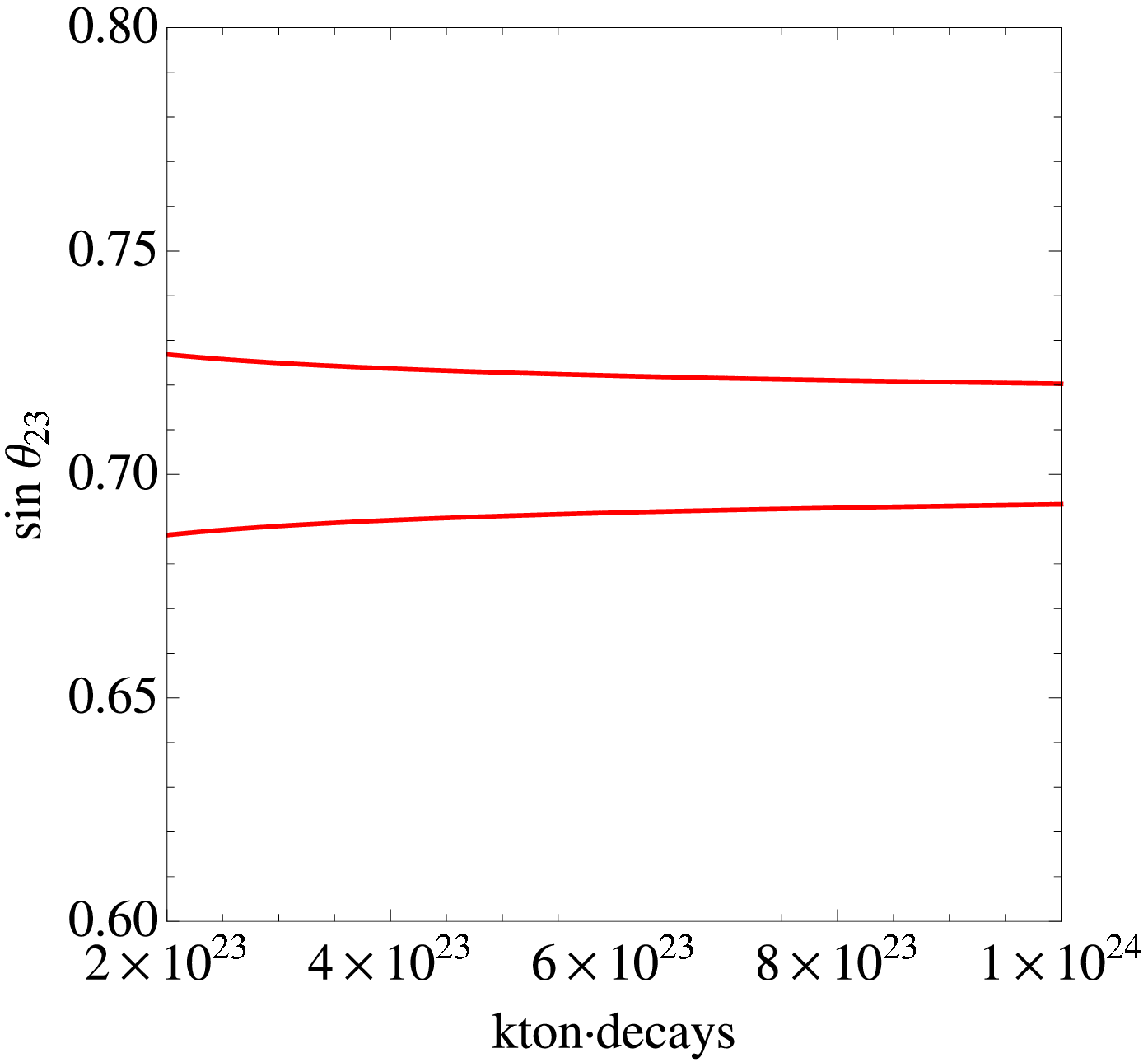}}\\
     \caption{$3\sigma$ regions for which maximal $\theta_{23}$ can be excluded, using a simulated value of $\theta_{13}=0^{\circ}$ when a) including systematic errors and backgrounds, b) no systematic errors and backgrounds are included.}
\label{fig:maxmix}
\end{figure}

We also explore the sensitivity to maximal mixing, i.e. the ability to exclude $\theta_{23}=45^\circ$, versus the exposure. We present the $3\sigma$ results in Fig.~\ref{fig:maxmix}. We have used a simulated value of $\theta_{13}=0^\circ$ here (so that $\delta$ is irrelevant) as the sensitivity to $\theta_{23}$ maximality comes primarily from the $\nu_{\mu}$ ($\bar{\nu}_{\mu}$) disappearance channels which are not dependent on $\theta_{13}$. Since the disappearance channels are also not strongly dependent upon systematic errors or backgrounds, there is little change obtained by switching these off.

\section{Liquid Argon Detector and comparison with other experiments}\label{sec:LAr}

Recently there has been much interest in the possibility of constructing a kton-scale liquid argon (LAr) detector~\cite{LAr}. If such a detector can be magnetised, it could be utilised in combination with a low energy neutrino factory and we have performed some preliminary studies to assess the potential of a 100 kton LAr detector for this experiment. As the design of large LAr detectors is still in the early stages, there are large uncertainties in the estimates for the detector performance. We assume an efficiency of $80\%$ on all channels and $5\%$ energy resolution for quasi-elastic events, then consider a range of values for other parameters. In the most conservative scenario, we assume $5\%$ systematics, $20\%$ energy resolution for non quasi-elastic events, and backgrounds of $5\times10^{-3}$ on the $\nu_{\mu}$ $(\bar{\nu}_{\mu})$ (dis)appearance channels and 0.8 on the $\nu_{e}$ $(\bar{\nu}_{e})$ appearance channels~\cite{Fleming}. For the optimistic scenario we use values identical to the TASD: $2\%$ systematics, $10\%$ energy resolution for non quasi-elastic events, and backgrounds of $1\times10^{-3}$ on the $\nu_{\mu}$ $(\bar{\nu}_{\mu})$ (dis)appearance channels and $1\times10^{-2}$ on the $\nu_{e}$ $(\bar{\nu}_{e})$ appearance channels. We find that varying the systematics, energy resolution and $\nu_{e}$ $(\bar{\nu}_{e})$ background do not play a large role in altering the results; the dominant effect comes from the variation of the $\nu_{\mu}$ $(\bar{\nu}_{\mu})$ background. 

\begin{figure}[!here]
     \centering
     \subfigure[~$\theta_{13}$ discovery potential]{
          \includegraphics[scale=0.5]{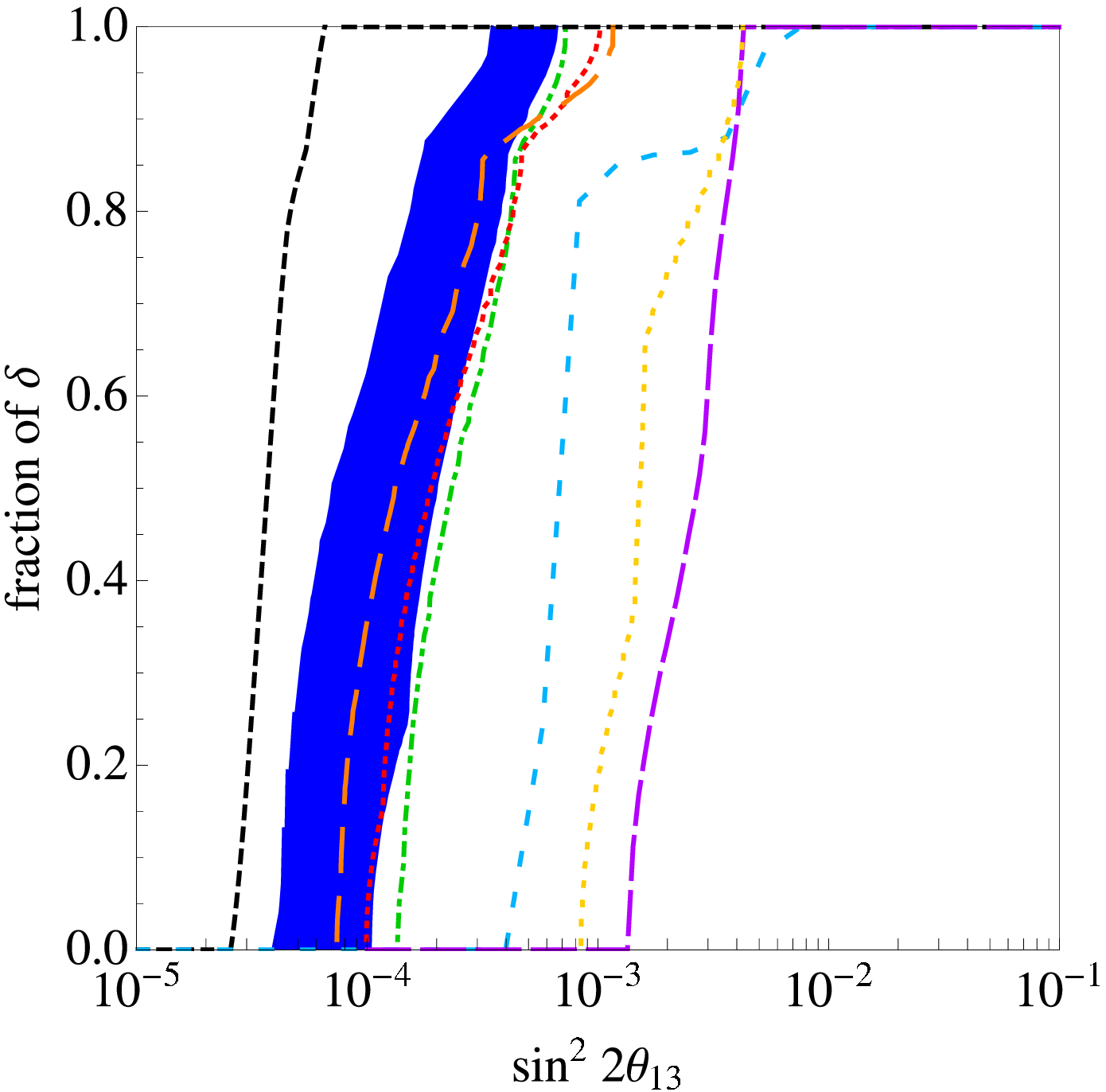}}
     \hspace{.3in}
     \subfigure[~CP discovery potential]{
          \includegraphics[scale=0.5]{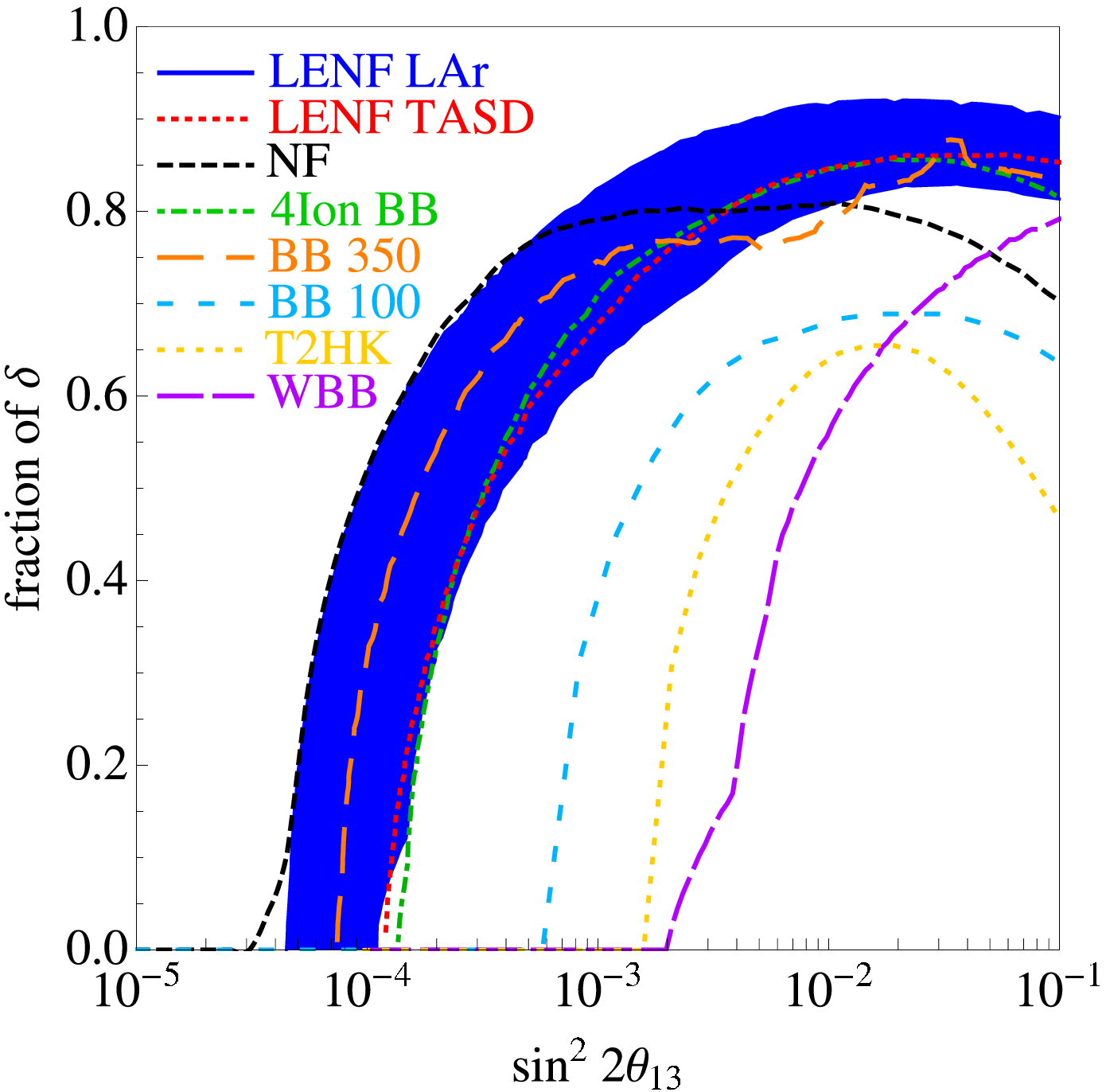}}\\
     \vspace{.3in}
     \subfigure[~Hierarchy sensitivity]{
          \includegraphics[scale=0.5]{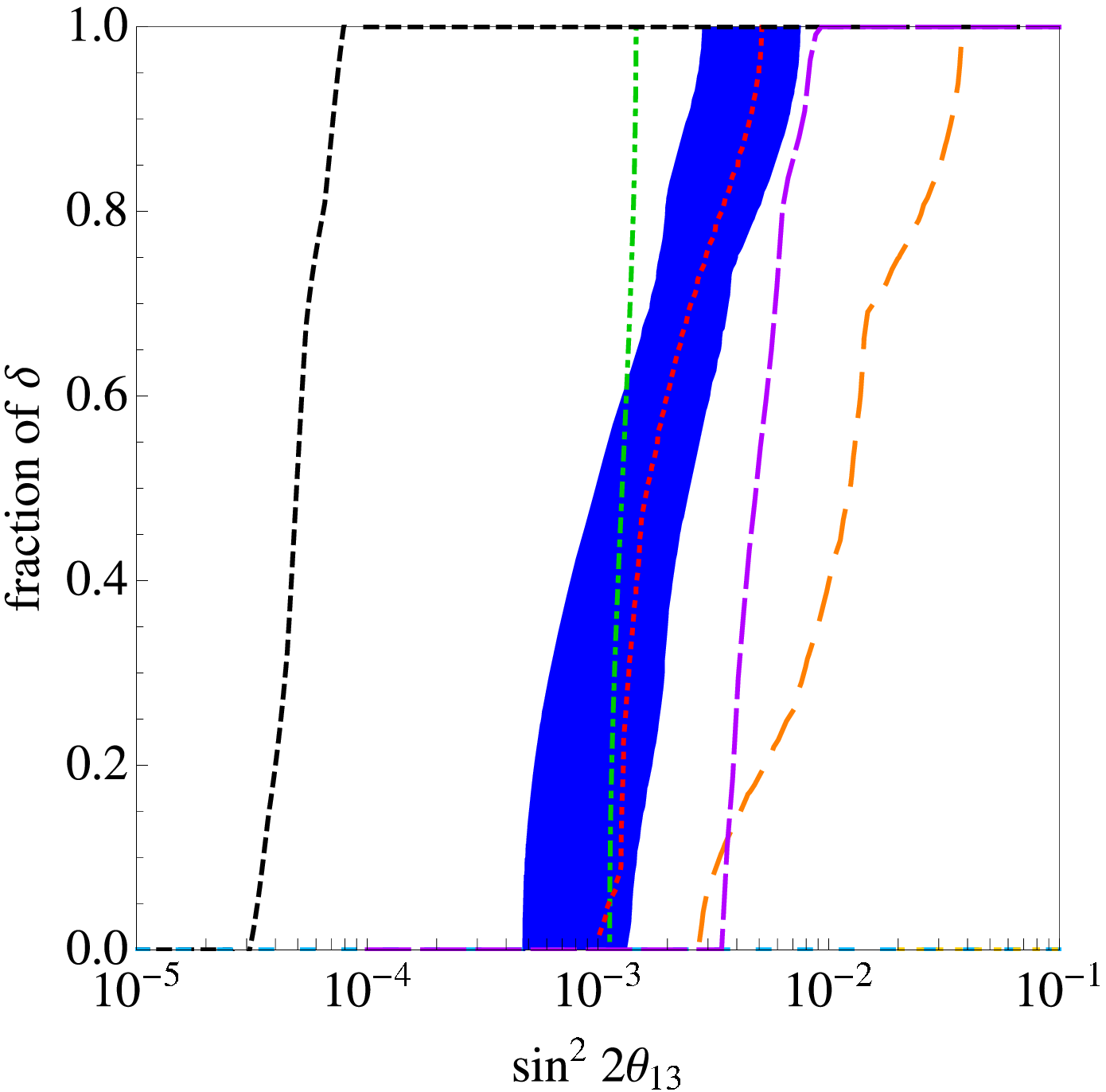}}
     \hspace{.3in}\\
     \caption{Comparison of $3\sigma$ allowed contours for the low energy neutrino factory with a 20 kton TASD (red line) and 100 kton LAr detector (blue band), the high energy neutrino factory (black line), the wide-band beam (purple line), T2HK (yellow line) and three $\beta$-beams (green, orange, light blue lines) for a) $\theta_{13}$ discovery potential, b) CP discovery potential, c) hierarchy sensitivity.}
\label{fig:exps}
\end{figure}

In Fig.~\ref{fig:exps} the results of the low energy neutrino factory with both the TASD and the two assumptions for the LAr detector are compared with other long-baseline experiments. We show the $3\sigma$ results for $\theta_{13}$ discovery, CP discovery potential, and hierarchy sensitivity (for normal hierarchy only) as a function of $\sin^{2}(2\theta_{13})$ in terms of the CP fraction. In order to make a fair comparison, we have used half the flux described in Section~\ref{sec:physics} for the low energy neutrino factory, to make it consistent with the other experiments which assume only $10^{7}$ seconds per year of observation. However, we believe that the fluxes in Section~\ref{sec:physics} are feasible. The results from the TASD are shown by the red line and those from the LAr detector are shown by the blue band. The right-hand edge of the band corresponds to the conservative estimate of the detector performance, and the left-hand edge to the most optimistic estimate. As the optimistic scenario assumes an almost identical performance to the TASD, the left-hand edge of the blue band also corresponds to the results obtainable from a 100 kton TASD. Results from the high energy neutrino factory~\cite{ISS-Detector Report}, wide-band beam~\cite{WBB}, T2HK~\cite{T2HK}, $100\gamma$ $ \beta$-beam~\cite{betabeam}, $350\gamma$ $\beta$-beam~\cite{gamma350} and 4-ion $\beta$-beam~\cite{4ion} are also shown.

In terms of sensitivity to $\theta_{13}$, a conservative low energy neutrino factory is an order of magnitude less sensitive than the high energy neutrino factory, but is still competitive with the $\beta$-beam experiments, giving an approximately equal performance to the 4-ion $\beta$-beam (which requires two baselines to resolve the degeneracy problem, as for the high energy neutrino factory). However, the performance of an aggressive low energy neutrino factory setup surpasses that of all other experiments except for the high energy neutrino factory. For CP violation, the low energy neutrino factory gives remarkable results: the most optimistic setup outperforms the high energy neutrino factory for all values of $\theta_{13}$, and even the most conservative setup gives a superior performance for $\sin^{2}(2\theta_{13}) > 2 \times 10^{-3}$, again equaling that of the 4-ion $\beta$-beam. For sensitivity to the mass hierarchy, the low energy neutrino factory gives an improvement over all other experiments apart from the higher energy setup and the 4-ion $\beta$-beam with their challenging $7000$ km baseline.

\section{Conclusions}\label{sec:conc}

We have optimised a low energy neutrino factory setup with a baseline of 1300 km, defining a reference setup to be one with a muon energy of 4.5 GeV and $1.4\times10^{21}$ useful muon decays per year, per polarity, running for ten years. For the detector we assume a totally active scintillating detector (TASD) with a fiducial mass of 20 kton, energy threshold of 0.5 GeV, energy resolution of $10\%$, efficiency for $\mu^{\pm}$ detection of $73\%$ below 1 GeV and $94\%$ above, efficiency for $e^{\pm}$ detection of $37\%$ below 1 GeV and $47\%$ above, and a background level of $10^{-3}$ on the $\nu_{e}\rightarrow\nu_{\mu}$ ($\bar{\nu}_{e}\rightarrow\bar{\nu}_{\mu}$) and $\nu_{\mu}\rightarrow\nu_{\mu}$ ($\bar{\nu}_{\mu}\rightarrow\bar{\nu}_{\mu}$) channels and $10^{-2}$ on the $\nu_{\mu}\rightarrow\nu_{e}$ ($\bar{\nu}_{\mu}\rightarrow\bar{\nu}_{e}$) channels. We have also considered a 100 kton liquid argon detector and found that its performance would equal or surpass that of the 20 kton TASD.

We have demonstrated how improving the energy resolution and statistics improves the performance of the setup, showing that in particular high statistics play a vital role. We have also shown how the combination of golden and platinum channels could be a powerful way of resolving degeneracies, especially in the case of limited statistics. However, once realistic background levels of at least $10^{-2}$ are considered, the improvement achieved by adding the platinum channel is negligible. Therefore, more effort should be invested into achieving larger statistics for the golden channel than in improving the platinum channel, at least for standard physics searches. 

Using our optimised setup, the low energy neutrino factory can have sensitivity to $\theta_{13}$ and $\delta$ for $\sin^{2}(2\theta_{13})>10^{-4}$, competitive with the high neutrino factory. Sensitivity to the mass hierarchy is accessible for $\sin^{2}(2\theta_{13})>10^{-3}$, better than other experiments using the same baseline due to the complementarity of measurements with different channels and different energies. Even if the flux is halved to equal that of other long-baseline experiments, the low energy neutrino factory is still competitive, performing especially well for CP discovery at large values of $\theta_{13}$. We have also studied the sensitivity to $\theta_{23}$, finding that it is possible to exclude maximal $\theta_{23}$ at $3\sigma$ for $\theta_{23}\lesssim43^{\circ}$ and $\theta_{23}\gtrsim47^{\circ}$, roughly independent of $\theta_{13}$, and to identify the octant for $\theta_{23}\lesssim37^{\circ}$ and $\theta_{23}\gtrsim53^{\circ}$.

Studies of the sensitivities as a function of exposure (detector mass $\times$ number of decays) show that the effect of non-zero systematic errors and backgrounds is to effectively halve the exposure, affecting the sensitivity to $\theta_{13}$, $\delta$ (especially for $\delta=0^{\circ}$) and $\theta_{23}$. For exposures $> 6\times10^{23}$ kton $\times$ decays per polarity and large $\theta_{13}$, the low energy neutrino factory could measure the oscillation parameters with unprecedented precision.

We conclude that the low energy neutrino factory has excellent sensitivity to the standard oscillation parameters and is therefore a potential candidate for a next-generation long-baseline experiment.

\vspace{1cm}

\section*{Acknowledgments}
 
This work was supported in part by the Fermi National Accelerator Laboratory, which is operated by the Fermi Research Association, under Contract No. DE-AC02-76CH03000 with the U.S. Department of Energy. SP and TL acknowledge the support of EuCARD (European Coordination for Accelerator Research and Development), which is co-funded by the European Commission within the Framework Programme 7 Capacities Specific Programme, under Grant Agreement number 227579. OM and SP would like to thank the Theoretical Physics Department at Fermilab for hospitality and support. TL also acknowledges the support of a STFC studentship and funding for overseas fieldwork. EFM acknowledges support by the DFG cluster of excellence `Origin and Structure of the Universe'. This work was undertaken with partial support from the European Community under the European Commission Framework Programme 7 Design Studies: EUROnu (Project Number 212372) and LAGUNA (Project Number 212343). The EC is not liable for any use that may be made of the information contained herein.

\pagebreak

\end{document}